\documentclass[prd,nofootinbib,preprint]{revtex4}
\pdfoutput=1
\usepackage{amsmath,amssymb,amsthm,amsxtra,overpic,bbm,bm,subfigure}
\usepackage{hyperref}
\usepackage{graphicx}
\usepackage{color}
\usepackage[utf8x]{inputenc}
\usepackage{array}
\newcommand{\PreserveBackslash}[1]{\let\temp=\\#1\let\\=\temp}
\newcolumntype{C}[1]{>{\PreserveBackslash\centering}p{#1}}
\newcolumntype{R}[1]{>{\PreserveBackslash\raggedleft}p{#1}}
\newcolumntype{L}[1]{>{\PreserveBackslash\raggedright}p{#1}}
\usepackage{comment}
\usepackage{epstopdf}
\usepackage{float}
\usepackage{multirow}
\usepackage{hyperref}
\usepackage[section]{placeins}
\usepackage{slashed,stmaryrd}
\usepackage{bbm}
\usepackage{lscape}%
\usepackage{array}
\usepackage{booktabs}%

\newcommand{\nua}[1]{\ensuremath{\rlap{\kern-2.5pt\ensuremath{\overset{\scriptscriptstyle(-)}{\phantom{\nu}}}}{\ensuremath{{\nu}_{#1}}}}\xspace}

\def\s1{\hat s}

\begin{document}
\title{ Linear seesaw in $A^\prime_5$ modular symmetry with Leptogenesis }

\author{ Mitesh Kumar Behera}
\email{miteshbehera1304@gmail.com}
\affiliation{School of Physics,  University of Hyderabad, Hyderabad - 500046,  India}

%\author{Subhasmita Mishra}
%\email{subhasmita.mishra92@gmail.com}
%\affiliation{School of Applied Sciences, Centurion University of Technology and Management, Odisha - 761211, India}
%
\author{ Rukmani Mohanta}
\email{rmsp@uohyd.ac.in}
\affiliation{School of Physics,  University of Hyderabad, Hyderabad - 500046,  India}

\begin{abstract}
In this paper, we investigate the implication of modular $\Gamma^{\prime}_5 \simeq A^{\prime}_5$ symmetry on  neutrino oscillation phenomenology in   the linear seesaw framework.   In order to achieve the  well defined mass structure for the light active neutrinos as dictated by the linear seesaw mechanism,  we introduce six heavy fermion fields along with a pair of weightons to retain the holomorphic nature of the superpotential.  The notable feature of modular symmetry is that, it reduces the usage of flavon fields significantly. In addition, the Yukawa couplings transform non-trivially  under the flavor symmetry group and expressed in terms of the Dedekind eta functions,  the $q$  expansion of which renders    numerical simplicity in calculations. We demonstrate that the model framework  diligently accommodates all the neutrino oscillation data.  Alongside, we also  investigate the effect of CP asymmetry generated from the decay of lightest heavy fermions to explain the  observed baryon asymmetry through the phenomenon of leptogenesis.
\end{abstract}

\maketitle
\flushbottom

\section{INTRODUCTION}
\label{sec:intro}
\label{sec:intro}

There are several unsolved knots in the realm of particle physics, e.g., the baryon asymmetry of the Universe, the dark matter content, the origin of neutrino  masses and mixing, etc.,  and the understanding of these issues  is one of the prime objectives of the present day research.  In the last couple of  decades, several diligent attempts have been made  towards comprehending and resolving the issue of  dynamical origin of fermion masses and their mixing. Present scenario has taken us few steps ahead in terms of getting a convincing explanation of the origin of mass through  Higgs mechanism while being within the domain of Standard Model (SM). However, it does not  provide proper grounds to explain the origin of the observed neutrino masses and their mixing. Rather, very diverse approaches are made in order to gain an insightful resolution towards the above existing problems, and obviously the answer lies in going beyond standard model (BSM) physics. It should be emphasized that, certain well-defined  patterns are observed in quark masses and mixing, the appreciation of which is still an enigma. Nonetheless, there are ample amount of  research work present, which make an attempt to grasp their fundamental origin. In addition, perplexity to the problem has increased due to the observation  of the neutrino masses and their sizeable mixing. The reason being, the order of magnitude of the observed neutrino masses  are approximately twelve order smaller than that of EW scale. Also, there is immense  difference in the pattern of leptonic and quark mixings, the former is  having large mixing angles, while the later involves smaller mixing angles.   Numerous experiments \cite{Bronner:2020eps,Pac:2018scx,Yu:2017woe,DoubleChooz:2019qbj} have confirmed the tininess of the neutrino mass and other parameters with high accuracy. The best-fit values of the neutrino oscillation parameters are furnished in Refs. \cite{deSalas:2020pgw, Esteban:2020cvm}. 

It is well-known that in the SM framework, the neutrino mass generation can not be explained through the standard Higgs mechanism  due to the  absence of the  right-handed (RH) components. Still, if we could manage to add the RH neutrinos into SM by hand,  and allow Dirac mass terms, the values of the required Yukawa couplings to be  around ${\mathcal O}(10^{-12})$,  which appear as aberrant. In contrast, there exist many BSM scenarios that help to generate tiny neutrino mass through the conventional seesaw mechanism. Some of the prominent seesaw mechanisms are categorized as type-I \cite{Mohapatra:1979ia,Brdar:2019iem,Branco:2020yvs,Bilenky:2010zza}, type-II \cite{Gu:2006wj,Luo:2007mq,Antusch:2004xy,Rodejohann:2004cg,Gu:2019ogb,McDonald:2007ka}, type-III \cite{Liao:2009nq,Ma:1998dn,Foot:1988aq,Dorsner:2006fx,Franceschini:2008pz,He:2009tf} and all of them require additional heavy fermions or scalars beyond the SM particle content. Literature survey shows there are many flavor symmetries either discrete $A_4$ \cite{King:2006np,Kalita:2015jaa,Altarelli:2007gb}, $S_3$ \cite{Kimura:2005sx,Mishra:2019keq,Meloni:2010aw,Pramanick:2019oxb}, $S_4$ \cite{Krishnan:2012me,Chakraborty:2020gqc,Vien:2016jkz} etc. or continuous $U(1)_{B-L}$ \cite{Ma:2014qra,Kanemura:2014rpa,Kanemura:2012zh,Mishra:2019gsr,Singirala:2017cch}, $U(1)_H$ \cite{Cai:2018upp,Nomura:2021adf,Dey:2019cts}, $U(1)_{L_e-L_\tau}$ \cite{Esmaili:2019pcy,Behera:2021nuk}  etc., which can  generate the tiny neutrino masses  and also accommodate  the observed neutrino oscillation data with the help of some additional scalars and perturbation (wherever required). As aforesaid, inclusion of flavons affects the neatness of the model and the predictability of the model is hampered because of the higher dimensional operators. These drawbacks can be eliminated through the recent approach of including modular symmetry \cite{Kobayashi:2018vbk,Feruglio:2017spp,deAdelhartToorop:2011re,Dudas:1995eq,Leontaris:1997vw,Du:2020ylx,Mishra:2020gxg,Okada:2019xqk,Penedo:2018nmg,Novichkov:2018ovf,Okada:2019lzv,Abbas:2020vuy,Nagao:2020snm,Asaka:2020tmo,Nomura:2020opk,Okada:2020dmb,Behera:2020lpd,Behera:2021eut,Behera:2020sfe,Ding:2019zxk,Altarelli:2005yx,Novichkov:2018nkm,Kashav:2021zir,Yao:2020zml}, where the  Yukawa couplings transform non-trivially under the discrete flavor symmetry group and  have certain modular weight.\\

The modular group $\Gamma^{\prime}_5 \simeq A^{\prime}_5$ is a new and promising candidate, which corresponds to the specific case of $N = 5$. People have done extensive studies on the basic properties of this finite group $A^\prime_5$ ~\cite{Everett:2010rd,Hashimoto:2011tn,Chen:2011dn}, so here we bring up only the important points regarding $A^\prime_5$ modular symmetry. The $A^\prime_{5}$ group consists  of 120 elements, which are likely to be produced by the generators $S$, $T$ and $R$ gratifying the identities for $N = 5$ \cite{Wang:2020lxk} . So, categorization of these 120 elements are done into nine conjugacy classes which are represented by  nine well defined irreducible representations, symbolized as ${\bf 1}$, $\widehat{\bf 2}$, $\widehat{\bf 2}^{\prime}_{}$, ${\bf 3}$, ${\bf 3}^{\prime}$, ${\bf 4}$, $\widehat{\bf 4}$, ${\bf 5}$ and $\widehat{\bf 6}$.  Additionally, the conjugacy classes and character table of $A^\prime_5$, as well as the representation matrices of all three generators $S$, $T$ and $R$, are presented in Appendix~\cite{Wang:2020lxk}.  It ought to be noticed that the  ${\bf 1}$, ${\bf 3}$, ${\bf 3}^{\prime}_{}$, ${\bf 4}$ and ${\bf 5}$ representations with $R = \mathbb{I}$ coincide with those for $A^{}_{5}$, while $\widehat{\bf 2}$, $\widehat{\bf 2}^{\prime}_{}$, $\widehat{\bf 4}$ and $\widehat{\bf 6}$ are unique for $A^\prime_5$ with $R = -\mathbb{I}$. As we are working in the modular space of $\Gamma(5)$, hence, its dimension is $5k+1$, where, $k$ is the modular weight.  A brief discussion concerning the modular space of $\Gamma(5)$ is presented in Appendix A.
For $k=1$, the modular space $M_1[\Gamma(5)]$ will have six basis vectors i.e., ($\widehat{e}_i$, where $i=1,2,3,4,5,6$) whose $q$-expansion are given below and they are used in expressing the Yukawa coupling $Y^{(1)}_{\widehat{\bf6}}$ as shown in Appendix B: 
\begin{eqnarray}
\widehat{e}^{}_1 & = & 1 + 3q + 4q^2_{} + 2q^3_{} + q^4_{} + 3q^5_{} + 6q^6_{} + 4q^7_{} - q^9_{} + \cdots \; , \nonumber \\
\widehat{e}^{}_2 & = & q^{1/5}_{} \left( 1 + 2q + 2q^2_{} + q^3_{} + 2q^4_{} + 2q^5_{} + 2q^6_{} + q^7_{} + 2q^8_{} + 2q^9_{} + \cdots \right), \nonumber \\
\widehat{e}^{}_3 & = & q^{2/5}_{} \left( 1 + q + q^2_{} + q^3_{} + 2q^4_{} + q^6_{} + q^7_{} + 2q^8_{} + q^9_{} + \cdots \right) ,\nonumber \\
\widehat{e}^{}_4 & = & q^{3/5}_{} \left( 1 + q^2_{} + q^3_{} + q^4_{} - q^5_{} + 2q^6_{} + 2q^8_{} + q^9_{} + \cdots \right) ,  \nonumber \\
\widehat{e}^{}_5 & = & q^{4/5}_{} \left( 1 - q + 2q^2_{} + 2q^6_{} - 2q^7_{} + 2q^8_{} + q^9_{} +\cdots \right) , \nonumber \\
\widehat{e}^{}_6 & = & q \left( 1 - 2q + 4q^2_{} - 3q^3_{} + q^4_{} + 2q^5_{} - 2q^6_{} + 3q^8_{} - 2q^9_{} + \cdots \right) ,
\label{eq:basisq}
\end{eqnarray} 
where $q \equiv e^{2 {\rm i} \pi \tau}$, with $\tau$ as a complex modulus parameter.
 The significance of the modulus $\tau$ is that, the modular group $\Gamma$ is generated by performing the linear  fractional transformation on $\tau $ as follows
\begin{equation}
\gamma :\tau \to \gamma(\tau) \rightarrow \frac{a \tau+b}{c \tau +d}\;,~~ \{a,b,c,d \in Z: ad-bc=0,~~{\rm Im}\tau >0\}.
\end{equation}
Our aim  here is to utilize the expediency of  $A^\prime_5$ modular symmetry by employing it to linear seesaw mechanism in the context of supersymmetry, as  we are quite familiar with the dynamics of  TeV scale seesaw  frameworks from numerous \cite{Grimus:2011mp,Ma:2009dk} literature. The deciding factor whether it will be linear seesaw or inverse seesaw is the position of the zero elements in the mass matrix under the basis of ($\nu$, $N_{R_i}$, $S_{L_i}$).  It is quite evident when 11 and 33 elements of the mass matrix are zero, it gives the structure of linear seesaw. As mentioned above, introduction of three left-handed neutral fermions superfields $S_{L_i}$ alongside  three-right handed ones $N_{R_i}$ $(i=1,2,3)$ validates and produces the neutrino mass matrix structure of linear seesaw  which is intricate enough, and has been studied in the context of discrete $A_4$ flavor symmetry in  \cite{Sruthilaya:2017mzt,Borah:2018nvu,Borah:2017dmk}. In this work,  we are implementing it under $A^\prime_5$ modular symmetry. For this purpose, the heavy fermions $S_{Li}$ \& $N_{Ri}$ are assigned as \textbf{$3^\prime$} under $A^\prime_5$ symmetry and modular form of the Yukawa couplings leads to a  constrained structure.  After that we perform the numerical analysis  to look for the region which is acceptable in order to fit the neutrino data. Hence, prediction for the neutrino sector is done after fixing the allowed parameter space. 
%\vspace{1mm}

Structure of this paper is as follows. In Sec. \ref{sec:linear} we discuss the layout of the familiar linear seesaw framework with  $A^\prime_5$ modular  symmetry and its alluring feature, which leads to simple mass structure for the charged-leptons as well as neutral leptons,  utilizing the product rules of $A^\prime_5$ symmetry. We thereafter,  briefly discuss the phenomena of light neutrino masses and mixing in this framework.  The numerical analysis  pertaining to  different observables of neutrino sector and the input model  parameters is presented in Sec. \ref{sec:results}. We also briefly comment  on the non-unitarity effect. The discussion on leptogenesis  within the context of the proposed model is furnished in Sec. \ref{sec:lepto} and  our results are summarized in Sec. \ref{sec:con}.
\section{MODEL FRAMEWORK}
\label{sec:linear}
%%%%%%%%%%%%%%%%%%%%%%%%%%%%%%%%%%
%%%%%%%%%%%%%%%%%%%%%%%%%%%%%%%%
\begin{center} 
\begin{table}[tbh!]
%\begin{table}[h!]%[tbh!]
%\begin{tiny}
\centering
\begin{tabular}{|c||c|c|c|c|c|c||c|c|c|c|c|}\hline\hline  
%  & \multicolumn{6}{c||}{Fermions} & \multicolumn{2}{c|}{Scalars}  \\ \hline \hline

\textbf{Fields}& ~$e^c_R$~& ~$\mu^c_R$~  & ~$\tau^c_R$~& ~${L}_L$~& ~$N^c_R$~& ~$S_L$~& ~${H}_{u,d}$~&~$\zeta$&~$\zeta^\prime$ \\ \hline 
%%%
%$SU(3)_C$ & $\bm{3}$  & $\bm{3}$ & $\bm{3}$ & $\bm{1}$ & $\bm{1}$ & $\bm{1}$ & $\bm{1}$ & $\bm{1}$ & $\bm{1}$   \\\hline 
$SU(2)_{L}$ & 1  & 1  & 1  & 2  & 1  & 1& 2   & 1 &1    \\\hline 
$U(1)_{Y}$   & 1 & 1 & 1 & $-\frac12$  & 0 & 0  & $\frac12, -\frac12$  & 0  &0 \\\hline
%$U(1)_{B-L}$   & $1$ & $1$ & $1$   &$-1$  & $-1$   &$2$ &$0$ &$-1$ \\\hline
$U(1)_{B-L}$   & 1 & 1 & 1   &$-1$  & 1   &0 &0 &1&-1 \\\hline
%$U(1)_X$   & $1$ & $1$ & $1$   &$\red{-1}$  & $1$  & $2$  & $0$  &$1$  \\\hline
$A^\prime_5$ & 1 & 1 & 1 & 3 & $3'$ & $3'$ & 1 & 1 &1  \\ \hline
%$k_I$ & $0$ & $0$ & $0$ & $0$ & $-2$ & $0$ & $0$ & $0$ & $-2$\\

$k_I$ & 1 & 3 & 5 & 1 & 1 & 4 & 0  & 1&1\\ 
\hline
\end{tabular}
\caption{The particle spectrum  and their charges under the symmetry groups $SU(2)_L\times U(1)_Y\times U(1)_{B-L} \times A^\prime_5$ while $k_I$ represents the modular weight.}
\label{tab:fields-linear}
\end{table}
\end{center} 
%%%%%%%%%%%%%%%%%%%%%%%%%%%%%%%%%%%%%%%%
Here we work on a model under linear seesaw scenario in the context of supersymmetry (SUSY), where Table  \ref{tab:fields-linear} provided below expresses the particle content and their respective group charges. For exploring neutrino sector beyond standard model (BSM), we extend it with the discrete $A^\prime_5$ modular symmetry and a local $U(1)_{B-L}$ gauge symmetry. However, the local  $U(1)_{B-L}$  becomes the auxillary symmetry which has been added to avoid certain undesirable terms in the superpotential. The advantage of working in BSM is that we can add right handed neutrinos and extra fields, and hence, here we have included  three extra right-handed SM singlet superfields ($N_{Ri}$), three left handed singlet superfields ($S_{Li}$) and a pair weightons ($\zeta, \zeta'$) in the particle gamut. 
The transformation of extra added superfields is taken as \textbf{$3^\prime$} under the $A^\prime_5$ modular group. The $A^\prime_5$ and $U(1)_{B-L}$ symmetries are broken at a very high scale, much greater than the scale of electroweak symmetry breaking \cite{Dawson:2017ksx}. Mass acquisition by the extra singlet superfield happens by allocating non-zero vacuum expectation values (VEVs) to the  weightons $\zeta$ and $\zeta'$. The modular weight  assigned to various particles is denoted by $k_I$. One of the significant point of introducing modular symmetry is the curtailment of flavon (weighton) fields, which  otherwise are traditionally required while working in BSM with discrete symmetries, since the Yukawa couplings have  non-trivial group transformation under $A^\prime_5$ modular symmetry group, and their transformation are present in \cite{Wang:2020lxk}.

The complete superpotential is given by
\begin{eqnarray}
\mathcal{W} &=&  A_{M_l}\left[({L}_L  l^c_R)_{\textbf{3}} Y^{k_Y}_{\textbf{3}}\right]H_d  + \mu H_u H_d  +  G_D \left[(L_L N^c_R)_{\textbf{5}} Y_{\bf 5}^{(2)}\right] H_u + \\ \nonumber
&& G_{LS}\left[(L_L S_L)_{\bf 4} H_u\sum\limits^{2}_{i=1} Y^{(6)}_{{\bf 4},i}\right] \frac{\zeta}{\Lambda}+ B_{M_{RS}}\left[({S}_L N^c_R)_{\textbf{5}} \sum\limits^{2}_{i=1}Y_{\textbf{5},i}^{(6)}\right] \zeta'. \label{super-p}
\end{eqnarray}
where, $A_{\mathcal{M}_l}= (\alpha_{\mathcal{M}_l}, \beta_{\mathcal{M}_l}, \gamma_{\mathcal{M}_l}$), $l^c_R = (e^c_R, \mu^c_R, \tau^c_R$), $k_Y = (2, 4, 6)$, and $G_D = \rm{diag} \{g_{D_1}, g_{D_2}, g_{D_3}\}$,  $G_{LS} = \rm{diag\{g_{LS_1},g_{LS_2}, g_{LS_3}\}}$, $B_{M_{RS}} = \rm{diag}\{\alpha_{RS_1}, \alpha_{RS_2}, \alpha_{RS_3}\}$ represent the coupling strengths of various interaction terms. 

%%%%%%%%%%%%%%%%%%%%%%%%%%%%%%%%%
\vspace{2cm}
\subsection{ Mass terms for the charged leptons  ($M_{\ell}$)}

In order to have a clear and simplified structure for charged lepton mass matrix, we consider the three families of left-handed lepton doublets ($L_{L}$) to transform as $\boldmath{3}$ under the $A^\prime_5$ symmetry with  $U(1)_{B-L}$ charge $-1$. The right-handed charged leptons $N_R^c$ transform as singlets under  $A^\prime_5$  symmetry and have  $U(1)_{B-L}$ charge $+1$. However,  ($e^c_R$, $\mu^c_R$, $\tau^c_R$) are given the modular weight as 1, 3, 5 respectively. The Higgsinos ${H}_{u,d}$ are given charges 0, 1 under the $U_{B-L}$ and $A^\prime_5$ symmetries  with zero modular weights. The VEVs of these  Higgsinos  ${H}_u$ and ${H}_d$ are given  as $v_u/\sqrt2$ and $v_d/\sqrt2$ respectively. Moreover, Higgsinos VEVs are associated to SM Higgs VEV as $v_H = \frac12 \sqrt{v^2_u + v^2_d}$ and the ratio of their VEVs is expressed as $\tan\beta= ({v_u}/{v_d})$ and we use its value to be 5 in our analysis. The relevant superpotential terms for charged leptons obtained from (\ref{super-p}) are given as
\begin{align}
 \mathcal{W}_{M_l}  
%&y_{\ell_{}} {L_L} H \ell_R  \nonumber \\
&= \alpha_{M_l}  \left[({L_L} e^c_R)_{\bf 3} Y_{\bf 3}^{(2)} \right] H_d +\beta_{M_l}  \left[({L_L} \mu^c_R)_{\bf 3} Y_{\bf 3}^{(4)} \right] H_d + \gamma_{M_l}  \left[({L_L} \tau^c_R)_{\bf 3} \left\lbrace \sum\limits_{i=1}^2Y_{{\bf 3},i}^{(6)} \right\rbrace \right] H_d
 %                   + {\rm H.c.}, 
 \label{Eq:yuk-MD} 
 \end{align}
Working under $A^\prime_5$ modular group, its Kronecker product (as provided in Appendix C), leaves us with a non diagonal charged lepton mass matrix after the spontaneous symmetry breaking. The mass matrix takes the form
\begin{align}
M_{l}&=\frac{v_d}{\sqrt2}
\left[\begin{matrix}
\left(Y^{(2)}_{\bf 3}\right)^{}_1 && \left(Y^{(4)}_{\bf 3}\right)^{}_1 && \left(\sum\limits_{i=1}^2Y^{(6)}_{{\bf 3},i}\right)_1  \\
\left(Y^{(2)}_{\bf 3}\right)^{}_3 &&\left(Y^{(4)}_{\bf 3}\right)^{}_3 &&\left(\sum\limits_{i=1}^2Y^{(6)}_{{\bf 3},i}\right)_3  \\
\left(Y^{(2)}_{\bf 3}\right)^{}_2 &&\left(Y^{(4)}_{\bf 3}\right)^{}_2 && \left(\sum\limits_{i=1}^2Y^{(6)}_{{\bf 3},i}\right)_2  \\
\end{matrix}\right]_{LR}      \cdot
\left[\begin{array}{ccc}
 \alpha_{M_l}  & 0 & 0 \\ 
0 & \beta_{M_l} & 0 \\ 
0 & 0 & \gamma_{M_l} \\ 
\end{array}\right]    .        
\label{Eq:Mell} 
\end{align}\\

The charged lepton mass matrix ${M}_l$ can be diagonalised by  the unitary matrix $U_l$, giving rise to the physical masses $m_e,~m_\mu$ and $m_\tau$ as  
\begin{equation} 
U^\dagger_l {M}_l {M}^\dagger_l U_l = {\rm diag}(m^2_e, m^2_\mu, m^2_\tau)\;. 
\end{equation}
In addition, it also satisfies the following identities, which will be used for numerical analysis in section \ref{sec:results}:
\begin{eqnarray}
	{\rm Tr} \left( {M}^{}_l {M}^{\dagger}_l \right) &=& m^{2}_{e} + m^{2}_{\mu} + m^{2}_{\tau} \; ,  \nonumber\\
	{\rm Det}\left( {M}^{}_l {M}^{\dagger}_l \right) &=& m^{2}_{e} m^{2}_{\mu} m^{2}_{\tau} \; , \nonumber\\
	\dfrac{1}{2}\left[{\rm Tr} \left({M}^{}_l {M}^{\dagger}_l\right)\right]^2_{} - \dfrac{1}{2}{\rm Tr}\left[ ({M}^{}_l {M}^{\dagger}_l)^2_{}\right] &= & m^{2}_{e}m^{2}_{\mu}+m^{2}_{\mu}m^{2}_{\tau}+m^{2}_{\tau}m^{2}_{e} \; . \label{eq:tr}
	%    (3.4-3.5)
	\end{eqnarray}

\subsection{Dirac and pseudo-Dirac mass terms for the light neutrinos}
%\vspace{3mm}
In addition to lepton doublets transformation, hitherto, the heavy fermion superfields, i.e., $N_{R_i}$  ($S_{L_i}$) transform as triplet $\boldmath{3^\prime}$ under $A_5$ modular group with  $U(1)_{B-L}$ charge of $-1$ (0) along with modular weight $1$ ($4$) respectively. As discussed in Ref.  \cite{Wang:2020lxk}, the choice of Yukawa couplings depends on the equation $k^{}_{Y} = k^{}_{I^{}_1} + k^{}_{I^{}_2} + \cdots + k^{}_{I^{}_n}$ where $k_Y$ is the modular weight of Yukawa couplings and $\Sigma^{I_n}_{i=1} k_{I_n}$ is sum of the modular weights of all other particles present in the definition of superpotential terms. These Yukawa couplings are expressed  in terms of Dedekind eta-function  $\eta(\tau)$, and thus have $q$-expansion forms, in order to avoid the complexity in calculations.
The relevant  superpotential term involving the active and right-handed neutrinos can be expressed as
\begin{align}
 \mathcal{W}_{D}  
%&y_{\ell_{}} \overline{L_L} H \ell_R  \nonumber \\
&= G_D   \left[({L_L} N^c_R)_{\bf 5} Y_{\bf 5}^{(2)} \right] H_u\;,
   %                 + {\rm H.c.},
                     \label{Eq:yuk-MD1} 
\end{align}
where, $G_D$ is the diagonal matrix containing the  free parameters  and the modular weight  of the Yukawa coupling is equal to the sum of the the modular weights of all other particles present in  (\ref{Eq:yuk-MD1}).   The choice of the Yukawa coupling is made based on the Kroncker product rules for $A^\prime_5$ modular symmetry such that superpotential remains singlet. The resulting Dirac neutrino mass matrix is found to be
\begin{align}
M_D&=\frac{v_u}{\sqrt30} G_D
\left[\begin{array}{ccc}
\sqrt3 (Y_\textbf{5}^{(2)})_1 & (Y_\textbf{5}^{(2)})_4 & (Y_\textbf{5}^{(2)})_3 \vspace{2mm}\\ 
 (Y_\textbf{5}^{(2)})_5 &~~ -\sqrt2  (Y_\textbf{5}^{(2)})_3 &~~ -\sqrt2  (Y_\textbf{5}^{(2)})_2 \vspace{2mm}\\ 
 (Y_\textbf{5}^{(2)})_2 &~~ -\sqrt2  (Y_\textbf{5}^{(2)})_5 &~~ -\sqrt2  (Y_\textbf{5}^{(2)})_4 \\ 
\end{array}\right]_{LR}.                   
\label{Eq:Mell} 
\end{align}
As the transformation of the   sterile fermion superfield  $S_L$ is  same as $N_R$ under $A^\prime_5$ modular symmetry,  it allows us to  define a pseudo-Dirac mass term for the light neutrinos and the corresponding interaction superpotential is expressed as
\begin{align}
 \mathcal{W}_{LS}  
%&y_{\ell_{}} {L_L} H \ell_R  \nonumber \\
&= G_{LS}   \left[({L_L} S_L)_{\bf 4} \sum\limits^{i=1}_2Y_{{\bf 4},i}^{(6)} \right] H_u \left(\frac{\zeta}{\Lambda}\right)\;,
  %                  + {\rm H.c.},
                     \label{Eq:yuk-MD} 
\end{align}
where, $G_{LS}$ is a diagonal matrix containing three  free parameters and the choice of Yukawa coupling depends upon the idea of keeping the superpotential singlet. Thus, we obtain  the structure for the pseudo-Dirac neutrino mass matrix of the form,
\begin{align}
M_{LS}&=\frac{v_u}{2\sqrt6}\left(\frac{v_\zeta}{\sqrt2 \Lambda}\right) G_{LS}
\left[\begin{array}{ccc}
0 & -\sqrt2 \left(\sum\limits_{i=1}^2 Y_{\textbf{4},i}^{(6)}\right)_3 &-\sqrt2 \left(\sum\limits_{i=1}^2 Y_{\textbf{4},i}^{(6)}\right)_2 \vspace{2mm}\\ 
\sqrt2 \left(\sum\limits_{i=1}^2 Y_{\textbf{4},i}^{(6)}\right)_4 &-  \left(\sum\limits_{i=1}^2 Y_{\textbf{4},i}^{(6)}\right)_2 &\left(\sum\limits_{i=1}^2 Y_{\textbf{4},i}^{(6)}\right)_1 \vspace{2mm}\\ 
\left(\sum\limits_{i=1}^2 Y_{\textbf{4},i}^{(6)}\right)_1 &\left(\sum\limits_{i=1}^2 Y_{\textbf{4},i}^{(6)}\right)_4 &-\left(\sum\limits_{i=1}^2 Y_{\textbf{4},i}^{(6)}\right)_3 \\ 
\end{array}\right]_{LR}.                 
\label{Eq:yuk-LS} 
\end{align}\\

\subsection{ Mixing between the heavy fermions $N_{Ri}$ and $S_{Li}$}
%\vspace{3mm}
Introduction of extra symmetries, helps in allowing the mixing of heavy superfields but forbids the usual Majorana mass terms. Hence, below we exhibit the mixing of these extra superfields i.e., $(N_{R}, S_{L})$ as follows
\begin{eqnarray}
 \mathcal{W}_{M_{RS}}  
                  =~ B_{M_{RS}}\left[({S_L} N^c_R)_{\bf 5} \sum\limits_{i=1}^2 Y_{{\bf 5},i}^{(6)}\right] \zeta',     
      \label{Eq:yuk-M} 
\end{eqnarray}
where, $B_{M_{RS}}$ is the free parameter \& $\langle \zeta' \rangle = v_{\zeta'}/\sqrt{2}$ is the VEV of $\zeta'$ and the superpotential is singlet under the $A^\prime_5$ modular symmetry product rule. Thus, considering $v_{\zeta'}\approx v_\zeta$, one can obtain the mass matrix  as follows:
\begin{align}
M_{RS}&=\frac{v_\zeta}{\sqrt60}B_{M_{RS}}
\left[\begin{array}{ccc}
2 \left(\sum\limits_{i=1}^2 Y_{\textbf{5},i}^{(6)}\right)_1 & -\sqrt3\left(\sum\limits_{i=1}^2 Y_{\textbf{5},i}^{(6)}\right)_4 & -\sqrt3\left(\sum\limits_{i=1}^2 Y_{\textbf{5},i}^{(6)}\right)_3 \vspace{2mm}\\ 
 -\sqrt3\left(\sum\limits_{i=1}^2 Y_{\textbf{5},i}^{(6)}\right)_4 &\sqrt6  \left(\sum\limits_{i=1}^2 Y_{\textbf{5},i}^{(6)}\right)_2 &-\left(\sum\limits_{i=1}^2 Y_{\textbf{5},i}^{(6)}\right)_1 \vspace{2mm}\\ 
-\sqrt3 \left(\sum\limits_{i=1}^2 Y_{\textbf{5},i}^{(6)}\right)_3 &-\left(\sum\limits_{i=1}^2 Y_{\textbf{5},i}^{(6)}\right)_1 &\sqrt6  \left(\sum\limits_{i=1}^2 Y_{\textbf{5},i}^{(6)}\right)_5 \\ 
\end{array}\right]_{LR}.  \label{yuk:MRS}
\end{align}
The masses for the heavy superfields can be found in the basis $( N_R, S_L)^T$ as
\begin{eqnarray}
M_{hf}= \begin{pmatrix}
0 & M_{RS}\\
M^T_{RS} & 0
\end{pmatrix}.\label{mrs matrix}
\end{eqnarray}
Hence, one can have three doubly degenerate mass eigenstates for the heavy superfields upon diagonalization.
\subsection{Linear Seesaw framework for the light neutrino masses}
In the present scenario of $A^\prime_5$ modular symmetry, the light neutrino masses can be generated in the framework of  linear seesaw.   Thus, the  mass matrix in the  flavor basis of $\left(\nu_L, N^c_R, S_L \right)^T$, can be manifested as 
\begin{eqnarray}
\mathbb{M} = \left(\begin{array}{c|ccc}   & \nu_L & N^c_R  & S_L   \\ \hline
\nu_L  ~&~ 0       ~&~ M_D       &~ M_{LS} \vspace{1mm}\\
N^c_R    ~&~ M^T_D         ~& 0       &~ M_{RS} \vspace{1mm}\\
S_L ~&~ M_{LS}^T     ~&~ M_{RS}^T    & 0
\end{array}
\right).
\label{eq:numatrix-complete}
\end{eqnarray}
The  mass formula for the light neutrinos in the framework of linear seesaw  is governed by the assumption that $ M_{RS} \gg M_D, M_{LS}$ and is given as
\begin{eqnarray}
m_\nu 
&=& M_D M_{RS}^{-1} M_{LS}^{T}+{\rm transpose}.\label{mass}
%\left(\frac{M_D}{M_{NS}}\right) \mu \left(\frac{M_D}{M_{NS}}\right)^T\,. 
\end{eqnarray}
Besides the light neutrino masses,  other related parameters in the leptonic sector are the Jarlskog invariant, which signifies the measure of CP violation  and the effective neutrino mass  parameter $m_{ee}$ that plays a key role in the  neutrinoless double beta decay process.   These parameters can be obtained from the  PMNS matrix elements  through the following relations:
\begin{eqnarray}
&& J_{CP} = \text{Im} [U_{e1} U_{\mu 2} U_{e 2}^* U_{\mu 1}^*] = s_{23} c_{23} s_{12} c_{12} s_{13} c^2_{13} \sin \delta_{CP},\\
&& m_{ee}=|m_{1} \cos^2\theta_{12} \cos^2\theta_{13}+ m_{2} \sin^2\theta_{12} \cos^2\theta_{13}e^{i\alpha_{21}}+  m_{3} \sin^2\theta_{13}e^{i(\alpha_{31}-2\delta_{CP})}|.
\end{eqnarray}
Tremendous experimental efforts are going on to measure the effective Majorana  mass parameter $ m_{ee}$ and it  is expected to be measured by KamLAND-Zen experiment in the near  future~\cite{KamLAND-Zen:2016pfg}.  

\section{NUMERICAL ANALYSIS}
\label{sec:results}
\begin{figure}[htpb]
\begin{center}
\includegraphics[height=50mm,width=75mm]{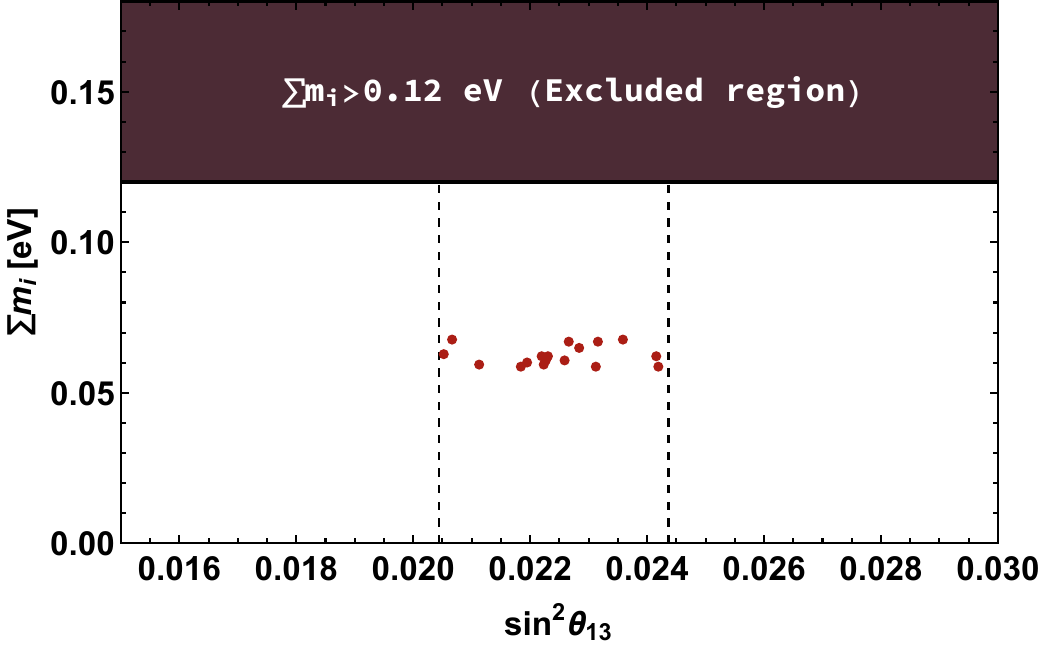}
\includegraphics[height=50mm,width=75mm]{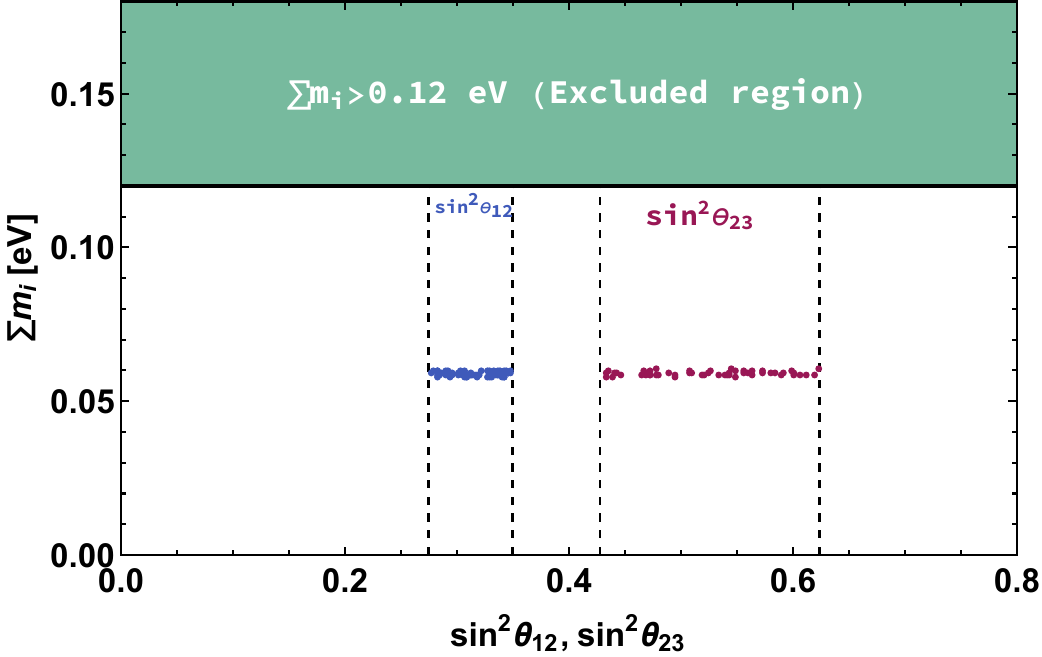}\\
\caption{ Left (Right) panel displays the correlation between $\sin^2 \theta_{13}$ ($\sin^2\theta_{12}$ \& $\sin^2 \theta_{23}$) with the sum of active neutrino masses. The vertical lines represent the $3 \sigma $ allowed ranges of the mixing angles.}
\label{mix_angles}
\end{center}
\end{figure}

\begin{table}[htb]
\centering
\begin{tabular}{|c|c|c|c|c|c|c|}
\hline
\bf{Oscillation Parameters} & \bf{Best fit} \bf{$\pm$ $1\sigma$} & \bf{ 2$\sigma$ range}& \bf{3$\sigma$ range} \\
\hline \hline
$\Delta m^2_{21}[10^{-5}~{\rm eV}^2]$& 7.56$\pm$0.19  & 7.20--7.95 & 7.05--8.14  \\
\hline
$|\Delta m^2_{31}|[10^{-3}~{\rm eV}^2]$ (NO) &  2.55$\pm$0.04 &  2.47--2.63 &  2.43--2.67\\
%$|\Delta m^2_{31}|[10^{-3}eV^2]$(IO)&  2.47$^{+0.04}_{-0.05}$ &  2.39--2.55 &  2.34--2.59 \\
\hline
$\sin^2\theta_{12} / 10^{-1}$ & 3.21$^{+0.18}_{-0.16}$ & 2.89--3.59 & 2.73--3.79\\
\hline
$\sin^2\theta_{23} / 10^{-1}$ (NO)
	  &	4.30$^{+0.20}_{-0.18}$ 
	& 3.98--4.78 \& 5.60--6.17 & 3.84--6.35 \\
%  $\sin^2\theta_{23} / 10^{-1}$ (IO)
	  & 5.98$^{+0.17}_{-0.15}$ 
	& 4.09--4.42 \& 5.61--6.27 & 3.89--4.88 \& 5.22--6.41 \\
\hline
$\sin^2\theta_{13} / 10^{-2}$ (NO) & 2.155$^{+0.090}_{-0.075}$ &  $1.98-2.31$ & $2.04-2.43$ \\
%$\sin^2\theta_{13} / 10^{-2}$ (IO) & 2.155$^{+0.076}_{-0.092}$ & 1.98--2.31 & 1.90--2.39 \\
\hline 
$\delta_{CP} / \pi$ (NO) & 1.08$^{+0.13}_{-0.12}$ & $0.84 - 1.42$ & $0.71 - 1.99$\\
\hline
\end{tabular}
\caption{The global-fit values of the  oscillation parameters alongwith their 1$\sigma$, 2$\sigma$ and 3$\sigma$ ranges \cite{deSalas:2017kay, Gariazzo:2018pei,Esteban:2018azc}.} \label{table:expt_value}
\end{table}
For numerical analysis, we use the neutrino oscillation parameters from the global-fit results \cite{deSalas:2017kay, Gariazzo:2018pei,Esteban:2018azc} obtained from various experiments, as given in Table \ref{table:expt_value}. The numerical diagonalization of the light neutrino mass matrix given in eqn.(\ref{mass}), is done through $U_\nu^\dagger {\mathcal M}U_\nu= {\rm diag}(m_1^2, m_2^2, m_3^2)$, where  ${\mathcal M}=m_\nu m_\nu^\dagger$ and $U_\nu$ is an unitary matrix. Thus, the lepton mixing matrix is given as  $U=U_l^\dagger U_\nu$, from which the mixing angles can be excerpted  using the standard relations:
\begin{eqnarray}
\sin^2 \theta_{13}= |U_{13}|^2,~~~~\sin^2 \theta_{12}= \frac{|U_{12}|^2}{1-|U_{13}|^2},~~~~~~~\sin^2 \theta_{23}= \frac{|U_{23}|^2}{1-|U_{13}|^2}\;.
\end{eqnarray}
To fit to the current neutrino oscillation data, we use the following ranges for  the model parameters:
\begin{align}
&{\rm Re}[\tau] \in [0,0.5],~~{\rm Im}[\tau]\in [1,3],~~G_{D} \in ~[10^{-7},10^{-6}],~~G_{LS} \in ~[10^{-4},10^{-3}] \quad v_\zeta  \in \nonumber [10,100] \ {\rm TeV}, \nonumber \\ &  B_{M_{RS}} \in ~[10^{-3},10^{-2}], \quad \Lambda \in  [10^4,10^5] \ {\rm TeV}. \label{input-values}
\end{align}
The input parameters are varied randomly in the ranges as provided in Eqn. (\ref{input-values}) and constrained by imposing the $3\sigma$ bounds on neutrino oscillation data,  i.e.,  the solar and atmospheric mass squared differences and the mixing angles as presented in Table \ref{table:expt_value}, as well as the  sum of active neutrino masses $\Sigma m_i < 0.12$ eV  \cite{Planck:2019nip,Planck:2018vyg}. The typical range of the modulus $\tau$ is found to be:  0\ $\lesssim\ $Re$[\tau]\lesssim$\ 0.5 and  1\ $\lesssim\ $Im$[\tau]\lesssim$\ 3 for normal ordered neutrino masses. In Fig. \ref{mix_angles}, we show the variation of the sum of active neutrino masses ($\Sigma m_i$) with the reactor  mixing angle $\sin^2 \theta_{13}$  in the left panel, while the right panel demonstrates $\Sigma m_i$ versus $\sin^2\theta_{12}$ and $\sin^2\theta_{23}$.  From these figures, it can be observed that the model predictions  for the sum of neutrino masses as $\Sigma 0.058~{\rm  eV} \leq m_i \leq 0.062$ eV for  the allowed $3 \sigma$ ranges of  the mixing angles. 
%%%%%%%%%%%%%%%%%%%%%%%%%%%%%%%%%%%%%%%%
%%%%%%%%%%%%%%%%%%%%%%%%%%%%%%%%%%%%%%%%%%%%%%

\begin{figure}[htpb]
\begin{center}
\includegraphics[height=50mm,width=75mm]{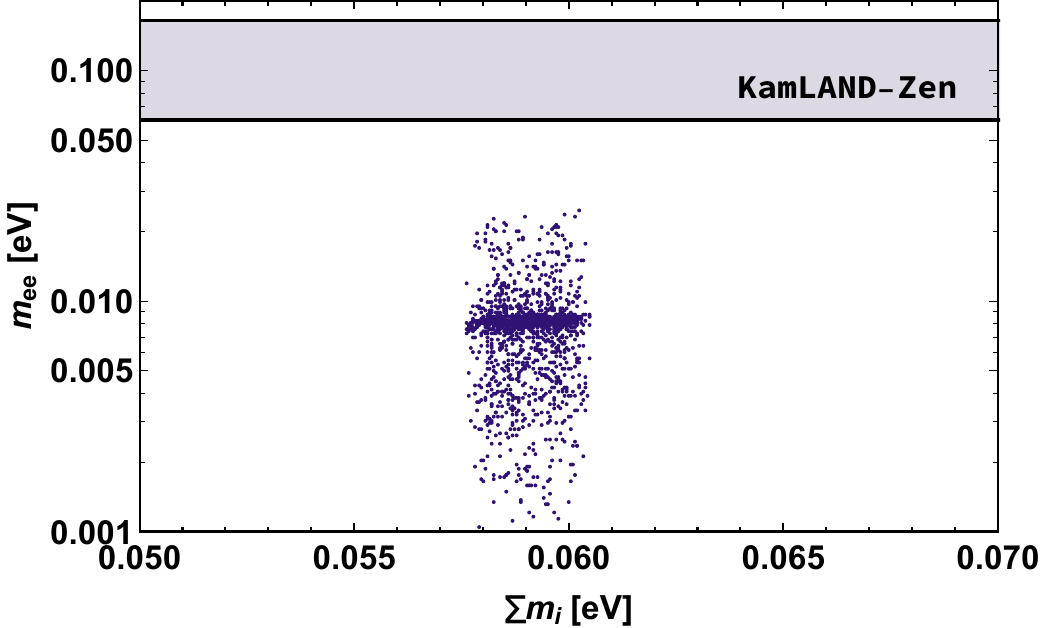}
%\hspace*{0.2 true cm}
%\includegraphics[height=48mm,width=72mm]{re_ss12_ss23.pdf}
\caption{Correlation plot between  the effective neutrino mass $m_{ee}$ of neutrinoless double beta decay and the sum of active neutrino masses.}
\label{M23_mee}
\end{center}
\end{figure}

\begin{figure}[htpb]
\begin{center}
\includegraphics[height=50mm,width=75mm]{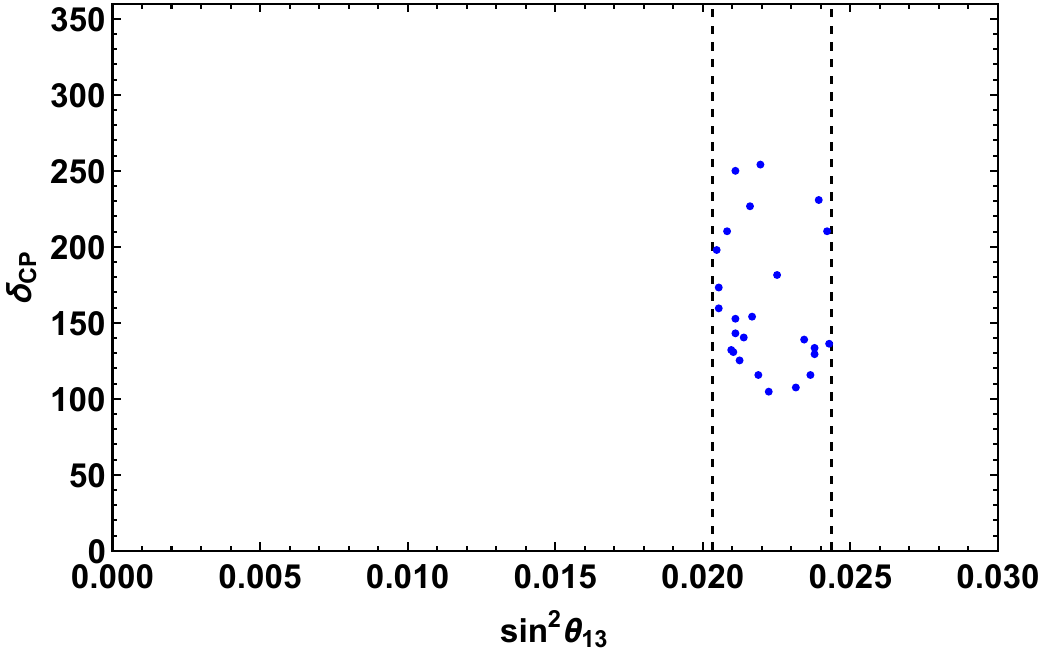}
\hspace*{0.2 true cm}
\includegraphics[height=48mm,width=72mm]{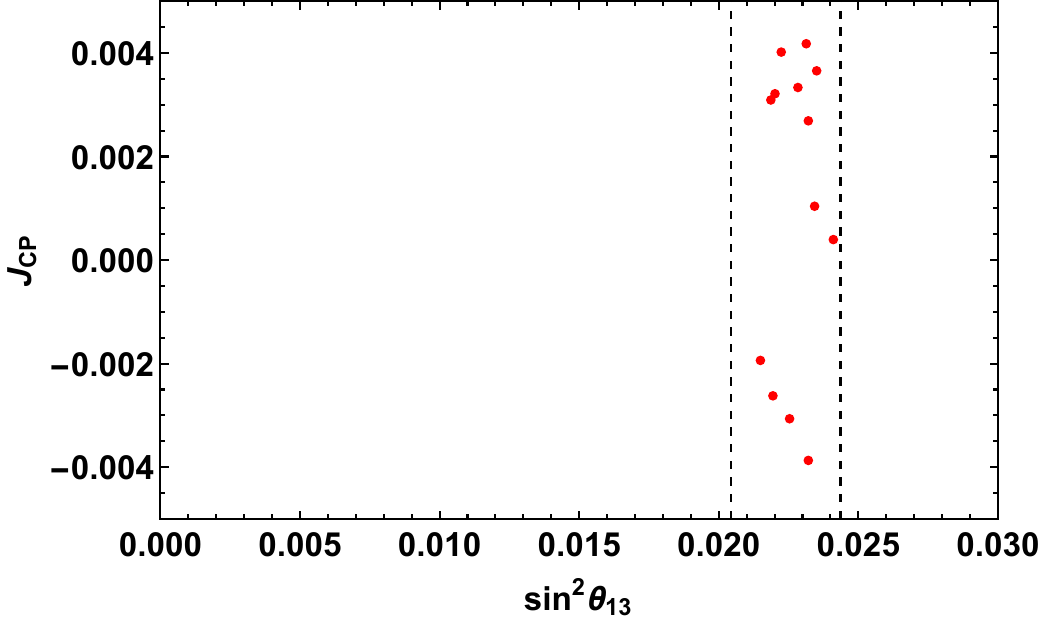}
\caption{Left (right) panel shows the plot of  $\delta_{CP}$ ($J_{CP}$) with $\sin^2 \theta_{13}$ within its $3\sigma$ bound.}
\label{dcpandjcpss13}
\end{center}
\end{figure}

The variation of the effective  neutrinoless double beta decay mass parameter $m_{ee}$ with $\Sigma m_i$ is displayed in  Fig. \ref{M23_mee},  from which the upper limit on $m_{ee}$ is found to be 0.025 eV satisfying KamLAND-Zen bound. Further, we display the variation of $\delta_{CP}$ and $J_{CP}$ in the left and right panel of Fig. \ref{dcpandjcpss13} respectively, where $100^\circ \leq \delta_{CP} \leq 250^\circ$ and $| J_{CP} | \leq 0.004$.
 
%%%%%%%%%%%%%%%%%%%%%%%%%%%%%%%%%%%%%%%%
%%%%%%%%%%%%%%%%%%%%%%%%%%%%%%%%%%%%%%%%
%\vspace{-5mm}
%%%%%%%%%%%%%%%%%%%%%%%%%%%%%%%%%%%%%%%%
%%\vspace{2cm}
\section*{Comment on non-unitarity of leptonic mixing matrix}
\label{sec:non-unitarity}
Here, we present a brief discussion on the non-unitarity of neutrino mixing matrix $U'_{\rm PMNS}$ in the context of the present model. Due to the mixing between the light and heavy fermions, there will be small  deviation from unitarity of the leptonic mixing matrix, which  can be expressed as following  \cite{Forero:2011pc}
\begin{align}
U'_{\rm PMNS}\equiv \left(1-\frac12 FF^\dagger\right) U_{\rm PMNS}.
\end{align}
Here,  $U_{\rm PMNS}$ denotes the leptonic mixing matrix that diagonalises the light neutrino mass matrix   and $F$ represents the mixing of active neutrinos with the heavy fermions,  approximated as $F\equiv  (M^{T}_{NS})^{-1} M_D \approx \frac{g_D v}{\alpha_{RS} v_\zeta}$, and is a hermitian matrix.
The local constraints on the non-unitarity parameters  \cite{Antusch:2014woa,Blennow:2016jkn,Fernandez-Martinez:2016lgt}, are obtained through various experimental results on electroweak parameters, e.g.,  the mass of $W$ boson ($M_W$),   the Weinberg mixing angle ($\theta_W$), several ratios of  fermionic   $Z$ boson decays as well as its  invisible decay, bounds from CKM unitarity, and lepton flavor violations. In the context of the present model, we presume the following approximated normalized order for the Dirac, pseudo-Dirac and heavy fermion masses for correctly generating  the observed solar and atmospheric mass-squared differences as well as the  sum of active neutrino masses of desired order as
\begin{eqnarray}
\left(\frac{m_\nu}{0.1 ~{\rm eV}} \right) \approx \left(\frac{M_D}{10^{-3}~~ {\rm GeV}}\right) \left(\frac{M_{RS}}{10^3~~ {\rm GeV}}\right)^{-1} \left( \frac{M_{LS}}{10^{-4}~~ {\rm GeV}}\right). 
\end{eqnarray}
With these chosen order masses, we obtain an approximated non-unitary mixing for the present model as
\begin{align}
|FF^\dagger|\le  
\left[\begin{array}{ccc} 
4.5\times 10^{-13}~~~~ &~~~~ 2.3\times 10^{-13}  ~~~~&~~ ~~6.2\times 10^{-13}  \\
2.3\times 10^{-13}  ~~~~&~~ ~~2.08\times 10^{-12} ~~~~ &~~~~ 4.5\times 10^{-12}  \\
6.2\times 10^{-13} ~~~~ & ~~~~4.5\times 10^{-12}  ~~~~&~~~~ 5.6\times 10^{-12} \\
 \end{array}\right].
\end{align} 
As the mixing between the active light and heavy fermions  in our model  is  quite small,  it  leads to a negligible contribution for the non-unitarity.
%%%%%%%%%%%%%%%%%%%%%%
%%%%%%%%%%%%%%%%%%%%%
%\begin{figure}[tb]\begin{center}
%\includegraphics[width=71mm, height=47mm]{y123_re.pdf}
%\hspace{6mm}
%\includegraphics[width=71mm, height=47mm]{y123_im.pdf}\\
%\caption{Here is the plots show the relation between the Yukawa couplings with Re $\tau$  in  (left panel) whereas (right panel) shows relation between Yukawa couplings with Im $\tau$.}   
%\label{fig:1}\end{center}\end{figure}
%%%%%%%%%%%%%%%%%%%
\section{Leptogenesis}
\label{sec:lepto}
The present universe  is clearly seen to be baryon dominated, with the ratio of the measured over-abundance of baryons over anti-baryons to the entropy density is found to be
\begin{equation}
Y_B = (8.56 \pm 0.22) \times 10^{-11}.
\end{equation} 
If the universe had started from an intially symmetric state of baryons and antibaryons, following three  conditions have to be fulfilled for generating a non-zero baryon asymmetry. According to   Sakharov \cite{Sakharov:1967dj}, these three criteria are: Baryon number violation, C and CP violation and departure from equilibrium during the evolution of the universe. Though the SM assures  all these criteria for an expanding Universe akin ours, the extent of CP violation found in the SM is quite  small to accommodate the observed baryon asymmetry of the universe. Therefore, additional sources of CP violation are absolutely essential for explaining this asymmetry. The most common  new sources of CP violation possibly could arise in the lepton sector, which is however,   not yet  firmly established. experimentally. Leptogenesis is the phenomenon that  furnishes a minimal set up to correlate the  CP violation in the lepton sector to the observed baryon asymmetry, as well as  imposes indirect constraints on the CP phases from the requirement that it would yield the correct baryon asymmetry. 
It is seen that the scale of  CP-asymmetry generated from the heavy neutral fermion decays can come down to  as low as TeV \cite{Pilaftsis:1997jf,Bambhaniya:2016rbb,Pilaftsis:2003gt, Abada:2018oly} due to resonant enhancement. However, the  present  scenario is realized by involving six heavy states, which comprises three pairs of heavy neutrinos with doubly degenerate masses  Eq.\eqref{mrs matrix}. Nevertheless, introduction of  a higher dimensional mass terms for the Majorana fermions ($N_R$) can be made through the following superpotential 
\begin{eqnarray}
\mathcal{W}_{M_R}=-G_R \left[\sum\limits_{i=1}^2Y^{(4)}_{{\bf 5},i} {N^c_R} N^c_R \right]\frac{{\zeta'}^2}{\Lambda}\;,
%+~{\rm H.c}.
\label{rhnmatrix}
\end{eqnarray} 
  which gives rise  to a petty mass splitting between the heavy neutral fermions, and hence provides an enhancement in the CP asymmetry for generating the required lepton asymmetry \cite{Pilaftsis:2005rv,Asaka:2018hyk}.
Thus, from (\ref{rhnmatrix})  one can construct the  Majorana mass matrix for the right-handed neutrinos $N_R$ as 
\begin{equation}
M_R =\frac{G_R v^2_\zeta}{2\Lambda\sqrt{30}}
\left[\begin{array}{ccc}
2 \left(\sum\limits_{i=1}^2 Y_{\textbf{5},i}^{(4)}\right)_1 & -\sqrt3\left(\sum\limits_{i=1}^2 Y_{\textbf{5},i}^{(4)}\right)_4 & -\sqrt3\left(\sum\limits_{i=1}^2 Y_{\textbf{5},i}^{(4)}\right)_3 \vspace{2mm}\\ 
 -\sqrt3\left(\sum\limits_{i=1}^2 Y_{\textbf{5},i}^{(4)}\right)_4 &\sqrt6  \left(\sum\limits_{i=1}^2 Y_{\textbf{5},i}^{(4)}\right)_2 &-\left(\sum\limits_{i=1}^2 Y_{\textbf{5},i}^{(4)}\right)_1 \vspace{2mm}\\ 
-\sqrt3 \left(\sum\limits_{i=1}^2 Y_{\textbf{5},i}^{(4)}\right)_3 &-\left(\sum\limits_{i=1}^2 Y_{\textbf{5},i}^{(4)}\right)_1 &\sqrt6  \left(\sum\limits_{i=1}^2 Y_{\textbf{5},i}^{(4)}\right)_5 \\ 
\end{array}\right]_{LR}.
\end{equation}
The coupling $G_R$ is considered as extremely small to preserve the linear seesaw texture of the neutrino mass matrix (\ref{eq:numatrix-complete}), i.e.,  $  M_D, M_{LS}\gg  M_R$ and hence,  inclusion of such additional term does not alter the previous results. However, this added term generates a small mass splitting.  Hence, the $2 \times 2$ submatrix of eqn. (\ref{eq:numatrix-complete}) in the  basis of $(N_R, S_L)$,  becomes
\begin{eqnarray}
M=\begin{pmatrix}
M_R& M_{RS}\\
M_{RS}^T & 0
\end{pmatrix},
\end{eqnarray}
which can be block diagonalized by the unitary matrix $ \frac{1}{\sqrt 2}
\begin{pmatrix}
I & -I\\
I & I
\end{pmatrix} $ as
\begin{eqnarray}
M'=\begin{pmatrix}
  M_{RS}+\frac{M_R}{2} & -\frac{M_R}{2}\\
-\frac{M_R}{2} &  -M_{RS}+\frac{M_R}{2}
\end{pmatrix} \approx  \begin{pmatrix}
  M_{RS}+\frac{M_R}{2} & 0\\
0 &  -M_{RS}+\frac{M_R}{2}
\end{pmatrix}.\label{block-diag}
\end{eqnarray}
Thus, one can express the mass eigenstates ($N^\pm$) in terms of the flavor states ($N_R, S_L$) as 
\begin{equation}
\begin{pmatrix}
S_{Li}\\N_{Ri}
\end{pmatrix}= \begin{pmatrix}
\cos{\theta} ~~&~ -\sin{\theta}\\
\sin{\theta}~ &~ \cos{\theta}
\end{pmatrix} \begin{pmatrix}
N_i^+ \\ N_i^-
\end{pmatrix}.
\end{equation}   
Assuming the mixing to be maximal, one can have
\begin{eqnarray}
N_{Ri} = \frac{(N_i^+ + N_i^-)}{\sqrt{2}},~~ S_{Li}= \frac{(N_i^+ - N_i^-)}{\sqrt{2}}.
\end{eqnarray}
Hence, the interaction superpotential (\ref{Eq:yuk-MD1}) can be manifested in terms of the new basis.  The mass eigenvalues of the new states $N^+$  and $N^-$ can be obtained by   diagonalizing the block-diagonal form of the  heavy-fermion masses and are found as $\frac{M_R}{2} + M_{RS}$ and $\frac{M_R}{2} - M_{RS}$ (\ref{block-diag}).

\begin{figure}[h!]
\begin{center}
\includegraphics[height=48mm,width=73mm]{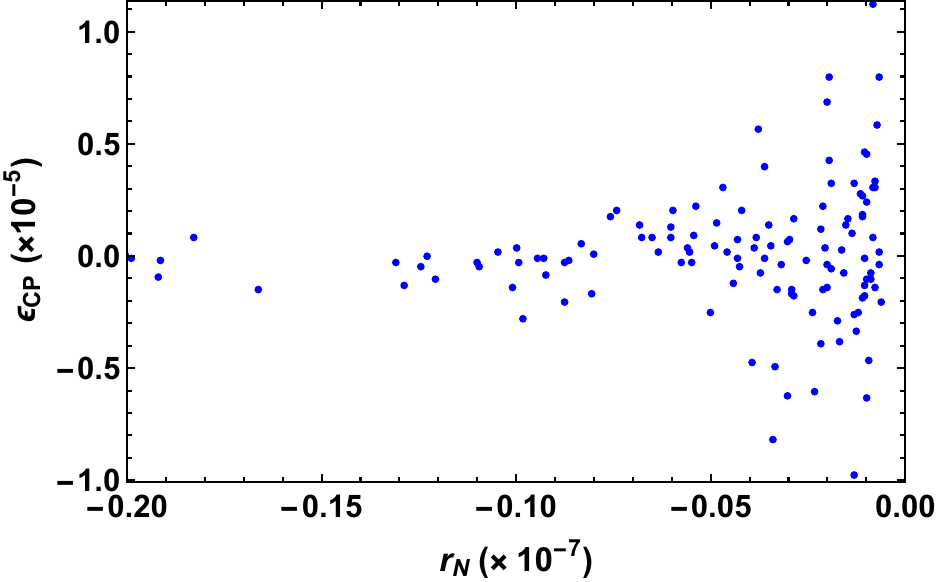}
\caption{Correlation  plot demonstrating the dependence of  CP asymmetry with the parameter $r_N$.}
\label{CP_var}
\end{center}
\end{figure}

The Dirac (\ref{Eq:yuk-MD}) and pseudo-Dirac (\ref{Eq:yuk-LS})   terms are now modified as
{\small{
\begin{eqnarray}
{\mathcal W}_D  =  G_{D} {L}_{L} {H_u} \left[Y^{(2)}_{\bf 5} \left(\frac{(N_i^+ + N_i^-)}{\sqrt{2}}\right)\right],
\end{eqnarray}}}
and
{\small{
\begin{eqnarray}
{\mathcal W}_{LS} =G_{LS}   {L}_{L} {H_u} \left[\sum\limits^{i=1}_2Y_{{\bf 4},i}^{(6)} \left(\frac{(N_i^+ - N_i^-)}{\sqrt{2}}\right)\right] \frac{\zeta}{\Lambda} .
\end{eqnarray}}}
Thus, one can symbolically  express the block-diagonal matrix for the heavy fermions (\ref{block-diag}) as
\begin{align}
M_{RS} \pm \frac{M_R}{2}&=\frac{v_\zeta}{\sqrt60}B_{M_{RS}} 
\begin{bmatrix}
2a~&~ d~&~e\\
d ~&~ b ~&~ f \\
e ~&~ f ~&~ c
\end{bmatrix} _{LR} \pm \frac{G_R v^2_\zeta}{2\Lambda\sqrt{30}}
\begin{bmatrix}
2a^\prime~&~ d^\prime~&~e^\prime\\
d^\prime ~&~ b^\prime ~&~ f^\prime \\
e^\prime ~&~ f^\prime ~&~ c^\prime
\end{bmatrix}_{LR},
\label{yuk:MRS}
\end{align} 
where, the different matrix elements are defined as 
\small{\begin{eqnarray}
a(a^\prime)=\left(\sum\limits_{i=1}^2 Y_{\textbf{5},i}^{6(4)}\right)_1,~ ~b(b^\prime)=\sqrt6  \left(\sum\limits_{i=1}^2 Y_{\textbf{5},i}^{6(4)}\right)_2,~ ~c(c^\prime)= \sqrt6  \left(\sum\limits_{i=1}^2 Y_{\textbf{5},i}^{6(4)}\right)_5 \;.\\
d(d^\prime)= -\sqrt3\left(\sum\limits_{i=1}^2 Y_{\textbf{5},i}^{6(4)}\right)_4, ~~e(e^\prime)= -\sqrt3\left(\sum\limits_{i=1}^2 Y_{\textbf{5},i}^{6(4)}\right)_3, ~~f(f^\prime)= -a(a^\prime)\;.
\end{eqnarray}}
One can obtain the diagonalized mass matrix from (\ref{yuk:MRS}) through rotation  to the mass eigen-basis as: $(M^{\pm})_{\rm diag}=U_{\rm TBM} U_R \left(M_{RS}\pm \frac{M_{R}}{2}\right) U^T_R U^T_{\rm TBM}$, and thus
  three sets of nearly degenerate mass states can be obtained  upon diagonalization.  We further presume that the lightest pair among them with mass in the TeV range,  contribute predominantly to the CP asymmetry. The small mass difference between the lightest pair demonstrates  that the CP asymmetry generated from the one-loop self energy contribution of heavy particle decay dominates over the vertex part.
Thus, the  CP asymmetry can be expressed as \cite{Pilaftsis:1997jf, Gu:2010xc}
\begin{eqnarray}
\epsilon_{{N_i^-}} 
\approx \frac{1}{32\pi^2 A_{N^-_i}}{\rm Im}\left[ \left(\frac{\tilde{M_D}}{v}-\frac{\tilde{M}_{LS}}{v}\right)^\dagger \left(\frac{\tilde{M}_D}{v}+\frac{\tilde{M}_{LS}}{v} \right)^2  \left(\frac{\tilde{M}_D}{v}-\frac{\tilde{M}_{LS}}{v}\right)^\dagger \right]_{ii} \frac{r_N}{r^2_N + 4 A^2_{N^-_i}}\;,\hspace*{0.5 true cm}
\end{eqnarray}
where $\tilde{M}_{D(LS)}= M_{D(LS)} U_{\rm TBM} U_R$, $\Delta M =M^+_i - M^-_i \approx M_R$ and the parameters $r_N$ and $A_{N^-}$ are given as
\begin{eqnarray}
&&  r_N =\frac{{(M^+_i)}^2 - {(M^-_i)}^2}{M^+_i M^-_i} = \frac{\Delta M (M^+_i + M^-_i)}{M^+_i M^-_i} ,\nonumber \\
&& A_{N^-} \approx \frac{1}{16\pi}\left[\left(\frac{\tilde{M}_D}{v}-\frac{\tilde{M}_{LS}}{v}\right)\left(\frac{\tilde{M}_D}{v}+\frac{\tilde{M}_{LS}}{v}\right) \right]_{ii}.
\end{eqnarray}
In Fig. \ref{CP_var}, we depict  the behavior of CP asymmetry with $r_N$, which satisfies both neutrino oscillation data and the CP asymmetry required for leptogenesis \cite{Davidson:2008bu,Buchmuller:2004nz}, which will be  discussed in the next subsection. 
%%%%%%%%%%%%%%%%%%%%%%%%%%%%%%%%
\subsection{Boltzmann Equations} 
Boltzmann equations are invoked to solve for the lepton asymmetry. It should be reiterated that, the Sakharov criteria \cite{Sakharov:1967dj} require the decay of the parent heavy fermion which ought to be out of equilibrium for generating the lepton asymmetry. In order to implement this condition, one has  to confront the Hubble rate to the decay rate as 
\begin{equation}
K_{N^-_i} = \frac{\Gamma_{N^-_i}}{H(T=M^-_i)}.
\end{equation}
Here, $H = \frac{1.67 \sqrt{g_\star}~ T^2 }{M_{\rm Pl}}$ is the Hubble expansion rate, with $g_{\star} = 106.75$ is the number of relativistic degrees  of  freedom in the thermal bath and $M_{\rm Pl} = 1.22 \times 10^{19}$ GeV is the Planck mass. Coupling strength becomes the deciding factor which guarantees inverse decay does not come into thermal equilibrium. For instance, if the strength is below $10^{-7}$ it gives $K_{N^-} \sim 1$. The Boltzmann equations associated with evolution of the number densities of right-handed fermion field and lepton, articulated in terms of yield parameter (ratio of number density to entropy density) are given by \cite{Plumacher:1996kc, Giudice:2003jh,Buchmuller:2004nz, Strumia:2006qk, Iso:2010mv}
\begin{eqnarray}
&& \frac{d Y_{N^-}}{dz}=-\frac{z}{s H(M_{N^-})} \left[\left( \frac{Y_{N^-}}{{Y^{\rm eq}_{N^-}}}-1\right)\gamma_D +\left( \left(\frac{{Y_{N^-}}}{{Y^{\rm eq}_{N^-}}}\right)^2-1\right)\gamma_S \right],\nonumber\\
&& \frac{d Y_{{B-L}}}{d z}= -\frac{z}{s H(M_{N^-})} \left[ 2\frac{Y_{{B-L}}}{{Y^{\rm eq}_{\ell}}}\gamma_{Ns}- \epsilon_{N^-} \left( \frac{Y_{N^-}}{{Y^{\rm eq}_{N^-}}}-1\right)\gamma_D \right],
\label{Boltz1}
\end{eqnarray}
where $s$ denotes the entropy density, $z = M^-_i/T$, $Y_{\mathcal{L}} = Y_{\ell} -Y_{\overline{\ell}}$ and the equilibrium number densities are given as \cite{Davidson:2008bu}
\begin{eqnarray}
%H(T)=\frac{4 {\pi}^3 g_\star}{45} \frac{T^2}{M_{\rm pl}},\hspace{3mm} %\hspace{3mm}  \text{where,} \hspace{3mm}    M_{\rm pl}=1.22\times 10^{19} ~\text{GeV},\\
Y^{\rm eq}_{N^-}= \frac{135  g_{N^-}}{16 {\pi}^4 g_\star} z^2 K_2(z), \hspace{3mm} {Y^{\rm eq}_\ell}= \frac{3}{4} \frac{45 \zeta(3) g_\ell}{2 {\pi}^4 g_{\star}}\,.
\end{eqnarray}
Here, $K_{1,2}$ are the modified Bessel functions, $g_\ell=2$ and $g_{N^-}=2$ represent the degrees of freedom of lepton and RH fermions, $\gamma_D$ is the decay rate and is given as
\begin{equation}
\gamma_D = s Y^{\rm eq}_{N^-}\Gamma_{N^-} \frac{K_1(z)}{K_2(z)}.
\end{equation}
While $\gamma_S$ represents the scattering rate of  $N^-N^- \to \zeta \zeta$ \cite{Iso:2010mv} and $\gamma_{Ns}$ denotes the scattering rate of $\Delta L = 2$ process. 
One can keep away the delicacy of the asymmetry being produced  even when $N^-$ is  in thermal equilibrium by subtracting the contribution arising  from on-shell $N^-$ exchange: ($\frac{\gamma_{D}}{4}$) from the total rate $\gamma_{Ns}$, given as  $\gamma^{\rm sub}_{Ns} = \gamma_{Ns} - \frac{\gamma_D}{4}$~ \cite{Giudice:2003jh}.  
\begin{figure}
\begin{center}
\includegraphics[height=65mm,width=95mm]{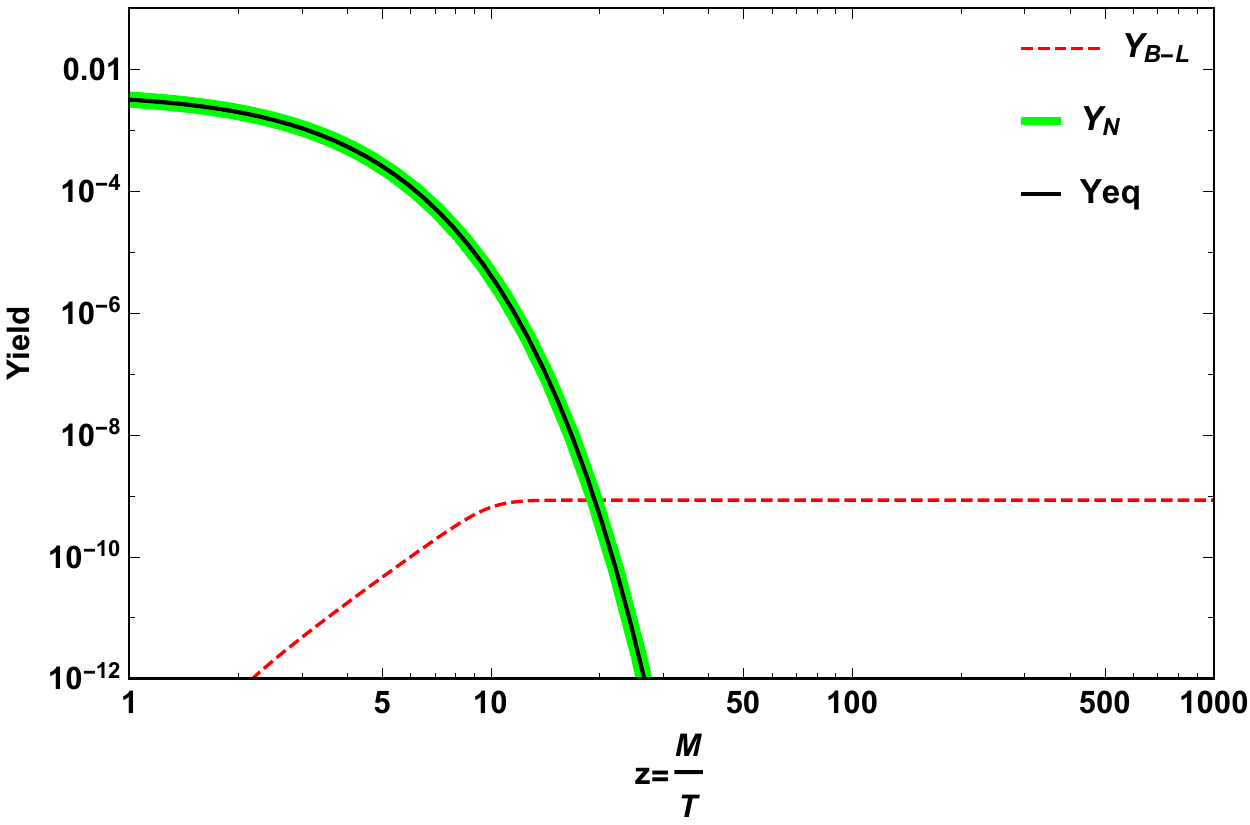}
\caption{Evolution of the yield parameters $Y_{N}$ and $Y_{\mathcal{L}}$ as a function of $z = M_{N^-}/{T}$.}\label{yield}
\end{center}
\end{figure}
The solution of Boltzmann eq. (\ref{Boltz1}) is displayed in Fig. \ref{yield}. For large coupling strength $Y_{N^-}$ (green-thick curve) almost traces $Y^{\rm eq}_{N^-}$ (black-solid curve) and the lepton asymmetry (red-dashed curve) is generated. The obtained lepton asymmetry can be converted to the  baryon asymmetry through the process of sphaleron transition, given as \cite{Plumacher:1996kc}
\begin{equation}
Y_B = -\left(\frac{8N_f + 4 N_H}{22 N_f + 13 N_H}\right)Y_{\mathcal{L}},
\end{equation}
where $N_f$ represents the number of fermion generations and $N_H$ denotes the number of Higgs doublets.
The observed baryon asymmetry can be expressed  in terms of baryon to photon ratio as \cite{Planck:2018vyg}
\begin{equation}
\eta = \frac{\eta_b - \eta_{\bar{b}}}{\eta_\gamma} = 6.08 \times 10^{-10}.
\end{equation}
The current bound on baryon asymmetry can be procured from the relation $Y_{B} = \eta/ 7.04$ as $Y_{B} \sim 8.6\times 10^{-11}$. Using the asymptotic value  of $Y_{\mathcal{L}}$ as ($8.77 \times 10^{-10}$)  from Fig. \ref{yield}, the obtained baryon asymmetry is $Y_{B} = -\frac{28}{79}~ Y_{\mathcal{L}} \sim 10^{-10}$.
%%%%%%%%%%%%%%%%%%%%%%%%%%%%%%%%%%
\subsection{A note on flavor consideration}
%%%%%%%%%%%%%%%%%%%%%%%%%%%%%%%%%
\begin{center} 
\begin{table}[htpb]
\centering
\begin{tabular}{|c|c|c|c|c|c|}  \hline
{$\epsilon^{e}_{N^-}$}  & ~{ $\epsilon^{\mu}_{N^-}$}~& ~$\epsilon^{\tau}_{N^-}$~ & $\epsilon_{N^-}$ & $\Delta M$~(GeV) \\\hline 
 $-1.78\times 10^{-5}$ & $-2.6 \times 10^{-5}$ & $-4.15 \times 10^{-5}$ &    $-8.53 \times 10^{-5}$ & $4\times 10^{-6}$  \\\hline
\end{tabular}
\caption{CP asymmetries and mass splitting obtained from the allowed range of model parameters which satisfy neutrino oscillation data.}
\label{tab:leptobench}
% \end{tiny}
\end{table}
\end{center}
%%%%%%%%%%%%%%%%%%%%%%%%%%%%%%%%%%%%%%
In leptogenesis, one flavor approximation is sufficient when  ($T>10^{12}$ GeV), meaning all the Yukawa interactions are out of equilibrium. But for temperatures $T\ll 10^{8}$ GeV, several charged lepton Yukawa couplings come into equilibrium making flavor effects an important consideration for generating the final lepton asymmetry. For temperatures below $10^6$ GeV, all the Yukawa interactions are in equilibrium and the asymmetry is stored in the individual lepton flavor. The detailed investigation of flavor effects in type-I leptogenesis can be seen in myriad literature \cite{Pascoli:2006ci,Antusch:2006cw,Nardi:2006fx,Abada:2006ea,Granelli:2020ysj,Dev:2017trv}. 

The Boltzmann equation for generating the lepton asymmetry in each flavor is \cite{Antusch:2006cw}
\begin{eqnarray}
\frac{d Y^{\alpha}_{ B-L_\alpha}}{d z}= -\frac{z}{s H(M_1^-)} \left[  \epsilon^\alpha_{N^-} \left( \frac{Y_{N^-}}{{Y^{eq}_{N^-}}}-1\right)\gamma_D-\left(\frac{\gamma^{\alpha}_D}{2}\right)\frac{A_{\alpha \alpha}Y^\alpha_{\rm B-L_\alpha}}{{Y^{eq}_{\ell}}}\right],
\end{eqnarray}
where, $\epsilon^\alpha_{N^-}$ i.e. $(\alpha=e,\mu,\tau)$ represents the CP asymmetry in each lepton flavor
\begin{equation}
\gamma_D^\alpha = s Y_{N^-}^{eq}\Gamma_{N^-}^\alpha \frac{K_1(z)}{K_2(z)}, \quad \gamma_D = \sum_\alpha \gamma^\alpha_D.\nonumber\\
\end{equation}
The matrix $A$ is given by \cite{Nardi:2006fx}, 
\begin{equation}
A=\begin{pmatrix}
-\frac{221}{711} && \frac{16}{711} && \frac{16}{711}\\
\frac{16}{711} && -\frac{221}{711} && \frac{16}{711}\\
\frac{16}{711} && \frac{16}{711}  && -\frac{221}{711} \\
\end{pmatrix}.\nonumber\\
\end{equation}

\begin{figure}
\begin{center}
\includegraphics[height=58mm,width=75mm]{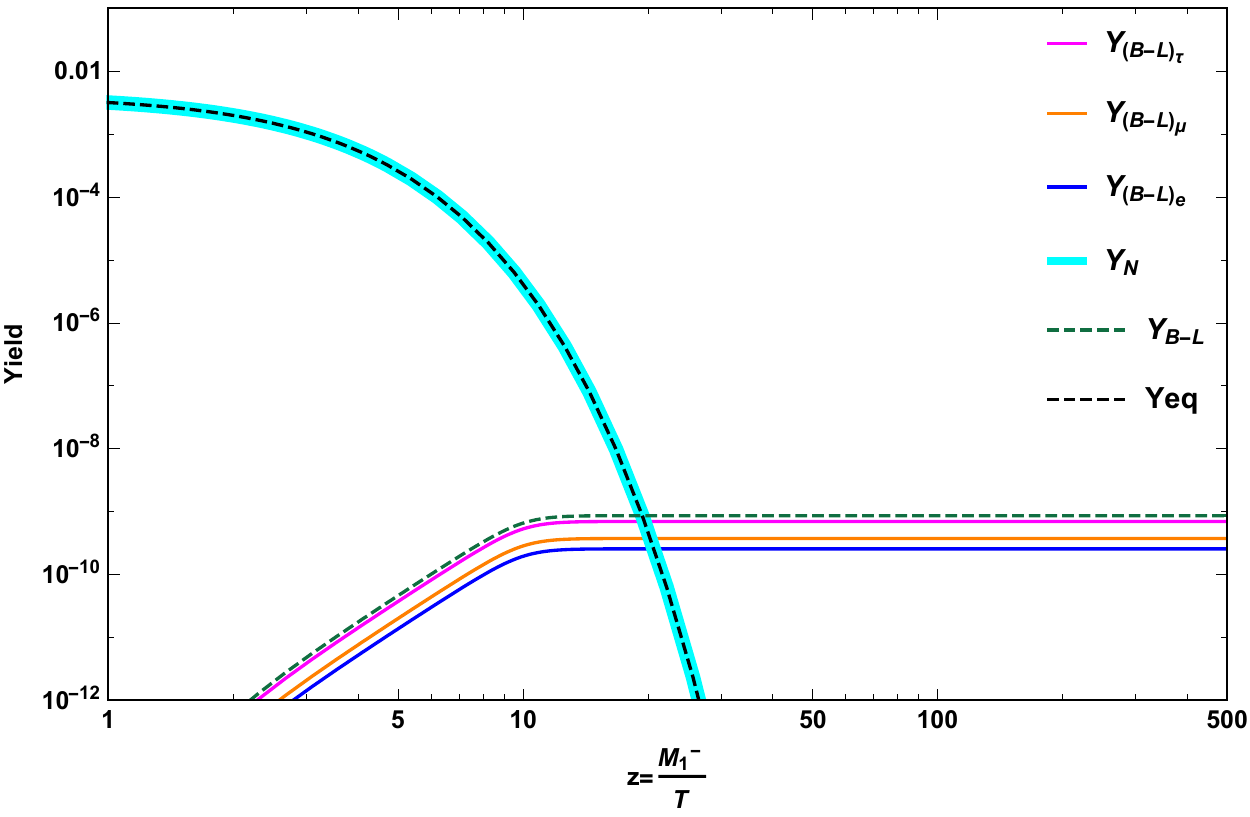}
\vspace{7mm}
\includegraphics[height=58mm,width=75mm]{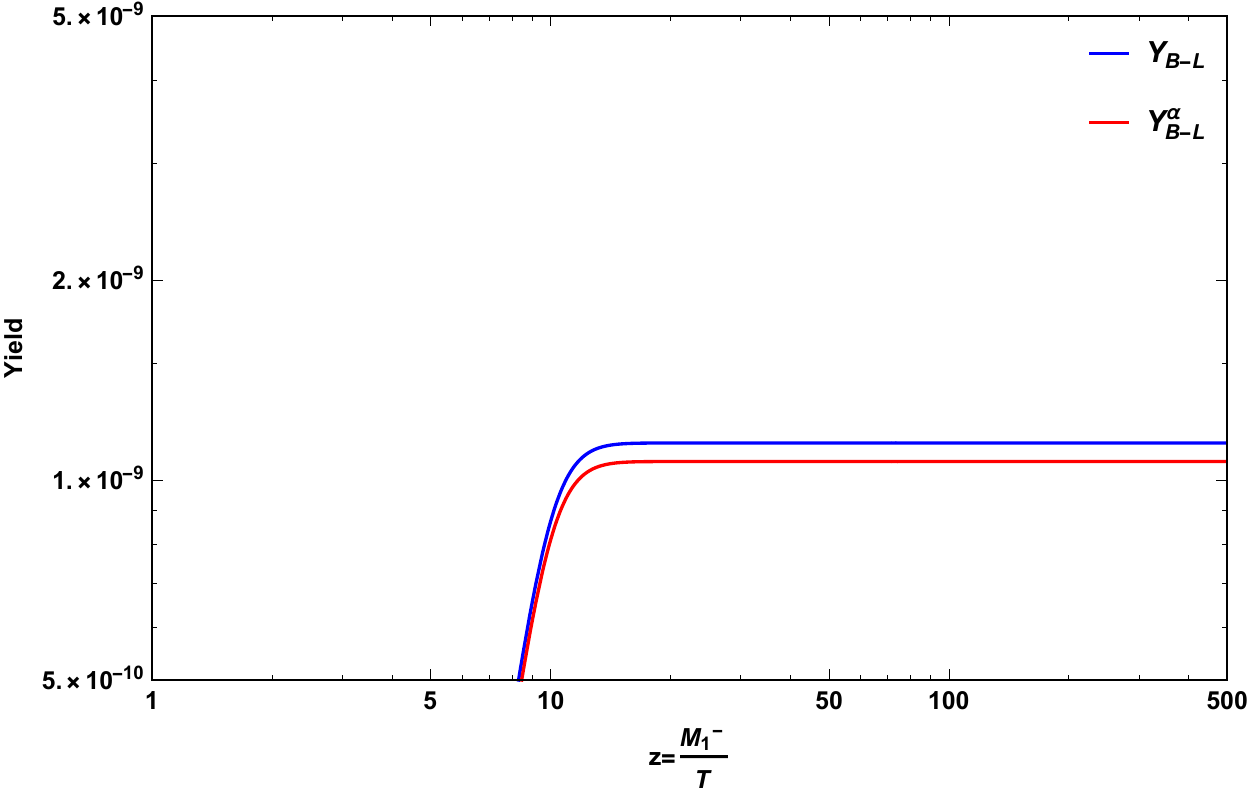}
\caption{After including the flavor effects the yield is shown in left panel, whereas, right panel shows the enhancement in the yield due to flavor effects.}\label{yield_Flavor}
\end{center}
\end{figure}

From the benchmark considered in Table. \ref{tab:leptobench}, we estimate the $B-L$ yield with flavor consideration in the left panel of Fig. \ref{yield_Flavor}. It is quite obvious to notice that the enhancement in $B-L$ asymmetry is obtained in case of flavor consideration (blue line) over the one flavor approximation (red line), as displayed in the right panel. This is because, in one flavor approximation the decay of the heavy fermion to a particular lepton flavor final state can get washed away by the inverse decays of any flavor unlike the flavoured case \cite{Abada:2006ea}.

%%%%%%%%%%%%%%%%%%%%%%%%%%%%%%%%%%%%%%%%%%%%%%%%%%%%%%
\section{Summary and Conclusion}
\label{sec:con}
In this work, we have investigated the implications of   $A^\prime_5$ modular symmetry on neutrino phenomenology. The important feature of the modular  flavor symmetry is that  it reduces the complications of accommodating multiple flavons, which are  usually associated with the use of discrete flavor symmetries. In the present model, we consider the SM to be extended by the  $A^\prime_5$ modular symmetry along with a $U(1)_{B-L}$ local gauge symmetry. It encompasses three right-handed as well as three left-handed heavy fermion fields to explore the neutrino phenomenology within the context of linear seesaw. In addition, it contains a pair of  singlet scalars, which  play a vital role in spontaneous breaking of $U(1)_{B-L}$ local symmetry and provide masses to the heavy fermions. The Yukawa couplings are considered  to   transform non-trivially under modular $A^\prime_5$ group, which basically replace the role of conventional flavon fields. This in turn, leads to a specific flavor structure for the neutrino mass matrix and helps  in exploring  the phenomenology of neutrino mixing. We numerically diagonalized the neutrino mass matrix to obtain the allowed regions for the model parameters, compatible with the current $3\sigma$ limit of oscillation data.  Further, our  model predicts the  CP violating phase $\delta_{CP}$ to be in the range of $(100^\circ-250^\circ)$ and the Jarlskog  invariant to be ${\mathcal O}(10^{-3})$.   The sum of active neutrino masses is found to be in the range  $0.058~\rm{eV} \leq \Sigma m_i \leq0.062~\rm{eV}$ and  the  value of effective neutrinoless double beta decay mass parameter $m_{ee}$ as  $(0.001-0.025)$ eV, which is quite below the current upper limits from KamLAND-Zen experiment i.e., $< (61-165)~ \rm{meV}$.  In addition, the flavor structure of heavy fermions gives rise to three sets of doubly degenerate mass eigenstates and hence, to incorporate leptogenesis, we introduced a higher dimension mass term for the right-handed neutrinos for generating a small mass splitting. We then obtained a non-zero CP asymmetry from the lightest heavy fermion decay  where the self energy contribution is partially enhanced due to the small mass splitting between the two lightest heavy fermions. Utilizing a particular benchmark of model parameters consistent with oscillation data, we tackled coupled Boltzmann equations to get the evolution of lepton asymmetry at TeV scale that emerges to be of the of the order $\simeq 10^{-10}$, which is adequate to explain the present baryon asymmetry of the universe. Besides, we have additionally shed light on the increase in asymmetry due to flavor consideration.
%%%%%%%%%%%%%%%%%%%%%%%%%%%%%%%%%%%%%%%%%%%%%

\acknowledgments

MKB acknowledges DST for its financial help. RM  would like acknowledge   SERB, Government of India for the support received through grant No. EMR/2017/001448 and University of Hyderabad IoE project grant no. RC1-20-012. We also acknowledge the use of CMSD HPC  facility of Univ. of Hyderabad for carrying out the computational work.

% BIBLIOGRAPHY
% use BIBTEX if you want
%\bibliographystyle{JHEP}
%\bibliography{yourBIBfiles}

\appendix
%\vspace{-1cm}
%\subsection{HIgher order Yukawa couplings}
%\label{HOYC}
\section*{Appendix A: The modular space of $\Gamma(5)$}
\label{mod_space}
In order to assert the modular forms which change nontrivially under $\Gamma^{\prime}_{5}$ $\simeq$ $A^\prime_5$, it is necessary to first find out the modular space of $\Gamma(5)$. Therefore, if $k$ is an integer i.e. non-negative, the modular space ${M}^{}_{k}\left[\Gamma(5)\right]$ bearing weight $k$ for $\Gamma(5)$ contains $5k+1$ linearly independent modular forms, which act like the basis vectors of the modular space. As stated in Ref.~\cite{Schultz:2015}, we have
\begin{eqnarray}
{M}^{}_k\left[\Gamma(5)\right] = \bigoplus^{}_{\substack{a+b = 5k \\a, b \geq 0}} \mathbb{C}\, \frac{\eta(5 \tau)^{15 k}}{\eta(\tau)^{3 k}} \, {\mathfrak k}^a_{\frac{1}{5},\frac{0}{5}}(5\tau) \, {\mathfrak k}^b_{\frac{2}{5},\frac{0}{5}}(5\tau) \; ,
\label{eq:G5basis}
%     (2.11)
\end{eqnarray}
represented below is the Dedekind eta function $\eta(\tau)$
\begin{eqnarray}
\eta(\tau)=q^{1 / 24}_{} \prod_{n=1}^{\infty}\left(1-q^{n}_{}\right) \; ,
\label{eq:Dedekindeta}
%     (2.12)
\end{eqnarray}
where, $q \equiv e^{2 {\rm i} \pi \tau}_{}$, and ${\mathfrak k}^{}_{r^{}_1,r^{}_2}(\tau)$ is the Klein form
\begin{eqnarray}
\mathfrak{k}^{}_{r^{}_{1}, r^{}_{2}}(\tau)= q_{z}^{(r^{}_{1}-1) / 2}\left(1-q^{}_{z}\right) \times \prod_{n=1}^{\infty}\left(1-q^{n}_{} q^{}_{z}\right)\left(1-q^{n}_{} q_{z}^{-1}\right)\left(1-q^{n}_{}\right)^{-2}_{} \; ,
\label{eq:Kexpansion}
%     (2.13)
\end{eqnarray}
where $(r^{}_1,r^{}_2)$ illustrates a pair of rational numbers in the domain of ${\mathbb Q}^2_{}-{\mathbb Z}^2_{}$, $z \equiv \tau r^{}_{1} + r^{}_{2}$ and $q^{}_z \equiv e^{2 {\rm i} \pi z}$. Under the transformations of $S$ and $T$, the eta function and the Klein form change as follows
\begin{eqnarray}
\begin{array}{cclcl}
S &: ~\quad & \eta(\tau) \rightarrow \sqrt{-{\rm i} \tau} \eta(\tau) \; , &~\quad &\mathfrak{k}^{}_{r^{}_{1}, r^{}_{2}}(\tau) \rightarrow -\dfrac{1}{\tau}\, \mathfrak{k}^{}_{-r^{}_{2}, r^{}_{1}}(\tau) \; , \\
T &: ~\quad &\eta(\tau) \rightarrow e^{\rm{i} \pi / 12}_{} \eta(\tau) \; , & ~\quad &\mathfrak{k}^{}_{r^{}_{1}, r^{}_{2}}(\tau) \rightarrow \mathfrak{k}^{}_{r^{}_{1}, r^{}_{1}+r^{}_{2}}(\tau) \; .
\end{array}
\label{eq:etaklein}
%     (2.14)
\end{eqnarray}
More information about the properties of the Kein form ${\mathfrak k}^{}_{r^{}_1, r^{}_2}(\tau)$ can be found in Refs.~\cite{Schultz:2015, Ding:2019xna}.
%%%%%%%%%%%%%%%%%%%%%%%%%%%%%%%%%%%%%%%%
\section*{Appendix B: Higher Order Yukawa couplings}
\label{HOYC}
All higher order Yukawa couplings are expressed in terms of the elements of $Y^{(1)}_{\widehat{\boldmath 6}}$ Yukawa coupling expressed as
%\vspace{-2cm}
\begin{eqnarray}
Y^{(1)}_{\widehat{\bf 6}} =
\left[\begin{matrix}
Y^{}_1 \\
Y^{}_2 \\
Y^{}_3 \\
Y^{}_4 \\
Y^{}_5 \\
Y^{}_6 \\
\end{matrix}\right]
=\left[\begin{matrix}
\widehat{e}^{}_{1}-3 \, \widehat{e}^{}_6 \\
5\sqrt{2} \, \widehat{e}^{}_2 \\
10 \, \widehat{e}^{}_3 \\
10 \, \widehat{e}^{}_4 \\
5\sqrt{2} \, \widehat{e}^{}_5 \\
-3 \, \widehat{e}^{}_1-\widehat{e}^{}_6 \\
\end{matrix}\right] \; .
\label{eq:Y_16}
%     (2.21)
\end{eqnarray}

Below are the modular form of the Yukawa couplings utilized to comstruct our model and the other couplings seen in the tensor product are expressed in \cite{Wang:2020lxk}
\begin{eqnarray}
Y^{(2)}_{{\bf 3}} &=& \left[Y^{(1)}_{\widehat{\bf 6}} \otimes Y^{(1)}_{\widehat{\bf 6}}\right]^{}_{{\bf 3}^{}_{{\rm s},1}} =
-3 \left[
\begin{array}{c}
\widehat{e}^{2}_1 - 36\, \widehat{e}_1 \widehat{e}^{}_6 - \widehat{e}_6^2 \\
5 \sqrt{2}\, \widehat{e}_2^{} (\widehat{e}^{}_1 - 3\, \widehat{e}_6^{}) \\
5 \sqrt{2}\, \widehat{e}_5^{} (3\, \widehat{e}^{}_1 + \widehat{e}_6^{}) \\
\end{array}
\right]
= -3 \left[
\begin{array}{c}
Y^2_1-3Y^{}_1Y^{}_6-Y^2_{6} \\
Y^{}_1 Y^{}_2 \\
-Y^{}_5 Y^{}_6 \\
\end{array}\right], 
\end{eqnarray}
\begin{eqnarray}
Y^{(4)}_{\bf 3} &=&  \left[Y^{(1)}_{\widehat{\bf 6}} \otimes Y^{(3)}_{\widehat{\bf 6},2}\right]^{}_{{\bf 3}^{}_{{\rm s},1}} = \frac{\sqrt{3}}{4}
\left[
\begin{array}{c}
\left(Y_1^2+Y_6^2\right)\left(7 Y_1^2-18  Y_1^{} Y_6^{}-7 Y_6^2\right) \\
Y_2^{} \left(13 Y_1^3-3 Y_1^{2} Y_6-29 Y_1^{} Y_6^2 -9
Y_6^3\right) \\
-Y_5^{} \left(9 Y_1^3-29 Y_1^2 Y_6^{} +3 Y_1^{} Y_6^2 +13
Y_6^3\right) \\
\end{array}
\right]  , 
\end{eqnarray}
\begin{eqnarray}
Y^{(6)}_{{\bf 3},1} &=& \left[ Y^{(1)}_{\widehat{\bf 6}} \otimes Y^{(5)}_{\widehat{\bf 2}^{\prime}_{}}\right]^{}_{\bf 3} =\dfrac{9\sqrt{2}}{16} \left(Y^2_1- 4 Y^{}_1 Y^{}_6 -Y^2_6\right) \left[
\begin{array}{c}
(Y_1-3 Y_6) (3 Y_1+Y_6) \left(3 Y_1^2-2 Y_1
Y_6-3 Y_6^2\right) \\
2 Y_2^{} \left(2 Y_1^3-9 Y_1 Y_6^2-3 Y_6^3\right) \\
2 Y_5^{} \left(3 Y_1^3-9 Y_1^2 Y_6+2 Y_6^3\right) \\
\end{array}
\right] \; , 
\label{eq:Y6_1}\nonumber\\
\end{eqnarray}
\begin{eqnarray}
Y^{(6)}_{{\bf 3},2} &=& \left[ Y^{(1)}_{\widehat{\bf 6}} \otimes Y^{(5)}_{\widehat{\bf 6},1}\right]^{}_{{\bf 3}^{}_{{\rm s},1}} =3\sqrt{2} \left(Y_1^4 - 3 Y_1^3 Y_6^{} - Y_1^2 Y_6^2 + 3 Y_1^{} Y_6^3 + Y_6^4\right) \left[
\begin{array}{c}
Y_1^2-3 Y_1^{} Y_6^{}-Y_6^2 \\
Y_1^{} Y_2^{} \\
-Y_5^{} Y_6^{} \\
\end{array}
\right] \; , 
\end{eqnarray}
\begin{eqnarray}
Y^{(6)}_{{\bf 4},1} &=& \left[ Y^{(1)}_{\widehat{\bf 6}} \otimes Y^{(5)}_{\widehat{\bf 2}}\right]^{}_{\bf 4} =-\dfrac{3}{4} \left(Y^2_1- 4 Y^{}_1 Y^{}_6 -Y^2_6\right)^2_{} \left[
\begin{array}{c}
-\sqrt{2} Y_2^{} (3 Y_1+Y_6) \\
Y_3^{} (Y_1+Y_6) \\
Y_4^{} (Y_1-Y_6) \\
\sqrt{2} Y_5^{} (Y_1-3 Y_6) \\
\end{array}
\right] \; , 
\end{eqnarray}
\begin{eqnarray}
Y^{(6)}_{{\bf 4},2} &=& \left[ Y^{(1)}_{\widehat{\bf 6}} \otimes Y^{(5)}_{\widehat{\bf 2}^{\prime}_{}}\right]^{}_{\bf 4} =-\dfrac{\sqrt{6}}{8} \left(Y^2_1- 4 Y^{}_1 Y^{}_6 -Y^2_6\right) \left[
\begin{array}{c}
\sqrt{2} Y_2^{} \left(Y_1^3+11 Y_1^2 Y_6^{}+19 Y_1^{}
Y_6^2+5 Y_6^3\right) \\
Y_3^{} \left(13 Y_1^3-31 Y_1^2 Y_6^{}-17 Y_1^{}
Y_6^2-Y_6^3\right) \\
Y_4^{} \left(Y_1^3-17 Y_1^2 Y_6^{}+31 Y_1^{} Y_6^2+13
Y_6^3\right) \\
\sqrt{2} Y_5^{} \left(5 Y_1^3-19 Y_1^2 Y_6^{} + 11 Y_1^{}
Y_6^2-Y_6^3\right) \\
\end{array}
\right]\; , 
\end{eqnarray}
\begin{eqnarray}
Y^{(2)}_{\bf 5} &=& \left[Y^{(1)}_{\widehat{\bf 6}} \otimes Y^{(1)}_{\widehat{\bf 6}}\right]^{}_{{\bf 5}^{}_{\rm s}}=
5\left[
\begin{array}{c}
 \sqrt{2} \left[\widehat{e}_1^2 + \widehat{e}_6^2\right) \\
-2 \sqrt{3}\, \widehat{e}_2^{} (\widehat{e}_1^{} + 7\, \widehat{e}_6^{}) \\
2 \sqrt{3}\, \widehat{e}_3^{} (4\, \widehat{e}_6^{} - 3\, \widehat{e}_1^{}) \\
-2 \sqrt{3}\, \widehat{e}_4^{} (4\, \widehat{e}_1^{} + 3\, \widehat{e}_6^{}) \\
2 \sqrt{3}\, \widehat{e}_5^{} (\widehat{e}_6^{} - 7\, \widehat{e}_1^{}) \\
\end{array}
\right]=\dfrac{1}{2}\left[
\begin{array}{c}
\sqrt{2}\left(Y^2_1+Y^2_6 \right) \\
2\sqrt{6}Y^{}_2 \left( 2Y^{}_1+Y^{}_6\right) \\
\sqrt{3}Y^{}_3\left( Y^{}_6-3Y^{}_1\right) \\
\sqrt{3}Y^{}_4 \left( Y^{}_1+3Y^{}_6\right) \\
2\sqrt{6}Y^{}_5 \left(2Y^{}_6-Y^{}_1\right) \\
\end{array}\right] \; .
\end{eqnarray}
\begin{eqnarray}
Y^{(4)}_{{\bf 5},1} &=&  \left[Y^{(1)}_{\widehat{\bf 6}} \otimes Y^{(3)}_{\widehat{\bf 4}}\right]^{}_{{\bf 5},2} =-\dfrac{3\sqrt{30}}{10} \left(Y^2_1 - 4 Y^{}_1 Y^{}_6 - Y^2_6 \right)
\left[
\begin{array}{c}
\sqrt{3} (Y_1^{}-3 Y_6^{}) (3 Y_1^{}+Y_6^{}) \\
-Y_2^{} (5 Y_1^{}+Y_6^{}) \\
\sqrt{2} Y_3^{} (Y_1^{}-Y_6^{}) \\
\sqrt{2} Y_4^{} (Y_1^{}+Y_6^{}) \\
-Y_5^{} (Y_1^{}-5 Y_6^{}) \\
\end{array}
\right] \; , 
\label{eq:Y4_1}
\end{eqnarray}
\\
\begin{eqnarray}
Y^{(4)}_{{\bf 5},2} &=&  \left[Y^{(1)}_{\widehat{\bf 6}} \otimes Y^{(3)}_{\widehat{\bf 6},2}\right]^{}_{{\bf 5}^{}_{{\rm a},1}} =  \dfrac{1}{4}
\left[
\begin{array}{c}
11 Y_1^4-60 Y_1^3 Y_6^{}+58 Y_1^2 Y_6^2+60 Y_1^{}
Y_6^3+11 Y_6^4 \\
\sqrt{3} Y_2^{} (Y_1^{}+Y_6^{}) \left(Y_1^2+8 Y_1^{} Y_6^{}+3
Y_6^2\right) \\
-\sqrt{6} Y^{}_3 \left(Y_1^3+3 Y_1^2 Y_6^{}-9 Y_1^{}
Y_6^2-3 Y_6^3\right) \\
\sqrt{6} Y_4^{} \left(3 Y_1^3-9 Y_1^2 Y_6^{}-3 Y_1^{}
Y_6^2+Y_6^3\right) \\
-\sqrt{3} Y_5^{} (Y_1^{}-Y_6^{}) \left(3 Y_1^2-8 Y_1^{}
Y_6^{}+Y_6^2\right) \\
\end{array}
\right] \; .
\label{eq:Y4_2}
%     (2.24)
\end{eqnarray}
\begin{eqnarray}
Y^{(6)}_{{\bf 5},1} &=& \left[ Y^{(1)}_{\widehat{\bf 6}} \otimes Y^{(5)}_{\widehat{\bf 4}}\right]^{}_{{\bf 5},2} =\dfrac{\sqrt{10}}{8} \left(Y^2_1- 4 Y^{}_1 Y^{}_6 -Y^2_6\right) \left[
\begin{array}{c}
\sqrt{3} (Y_1^{}-3 Y_6^{}) (3 Y_1^{}+Y_6^{})
\left(Y_1^2+Y_6^2\right) \\
-2 Y_2^{} (2 Y_1^{}+Y_6^{}) \left(2 Y_1^2-3 Y_1^{}
Y_6^{}-Y_6^2\right) \\
\sqrt{2} Y_3^{} \left(Y_1^3+2 Y_1^2 Y_6^{}-11 Y_1^{}
Y_6^2-4 Y_6^3\right) \\
\sqrt{2} Y_4^{} \left(4 Y_1^3-11 Y_1^2 Y_6^{}-2 Y_1^{}
Y_6^2+Y_6^3\right) \\
2 Y_5^{} (Y_1^{}-2 Y_6^{}) \left(Y_1^2-3 Y_1^{} Y_6^{}-2
Y_6^2\right) \\
\end{array}
\right] \; , 
\end{eqnarray}
\begin{eqnarray}
Y^{(6)}_{{\bf 5},2} &=& \left[ Y^{(1)}_{\widehat{\bf 6}} \otimes Y^{(5)}_{\widehat{\bf 6},1}\right]^{}_{{\bf 5}^{}_{\rm s}} =-\dfrac{1}{\sqrt{2}}\left(Y_1^4 - 3 Y_1^3 Y_6^{} - Y_1^2 Y_6^2 + 3 Y_1^{} Y_6^3 + Y_6^4\right) \left[
\begin{array}{c}
\sqrt{2} \left(Y_1^2+Y_6^2\right) \\
2 \sqrt{6} Y_2^{} (2 Y_1^{}+Y_6^{}) \\
-\sqrt{3} Y_3^{} (3 Y_1^{}-Y_6^{}) \\
\sqrt{3} Y_4^{} (Y_1^{}+3 Y_6^{}) \\
-2 \sqrt{6} Y_5^{} (Y_1^{}-2 Y_6^{}) \\
\end{array}
\right] \; . 
\end{eqnarray}
\section{$A_5^\prime$ Modular Symmetry Product Rule}
\label{app:B}
In this Appendix, we summarize the decomposition rules of the Kronecker products of any two nontrivial irreducible representations of $A^{\prime}_5$, namely, \vspace{3mm}\\
\renewcommand\arraystretch{0.9}
\vspace{1.5cm}
\noindent
\begin{small}
	\begin{tabular}{L{6cm}L{6cm}L{6cm}}
		\hspace{2mm}$\mathbf{3} \otimes \boldsymbol{3}=\mathbf{1}^{}_{\rm s} \oplus \boldsymbol{3}^{}_{\rm a} \oplus \boldsymbol{5}^{}_{\rm s}$ & \hspace{2mm}  ${\bf 3}^{\prime}_{} \otimes {\bf 3}^{\prime}_{} = {\bf 1}^{}_{\rm s} \oplus {\bf 3}^{\prime}_{\rm a} \oplus {\bf 5}^{}_{\rm s}$ &\hspace{2mm}  $ {\bf 3} \otimes {\bf 3}^{\prime}_{} = {\bf 4} \oplus {\bf 5}$ \\
		$\left\{\begin{array}{l}
		\mathbf{1}^{}_{\rm s} : \dfrac{\sqrt{3}}{3} (\varrho^{}_{1} \vartheta^{}_{1}+\varrho^{}_{2} \vartheta^{}_{3}+\varrho^{}_{3} \vartheta^{}_{2})\\
		{\bf 3}^{}_{\rm a} : \dfrac{\sqrt{2}}{2} \left(\begin{array}{c}
		\varrho^{}_{2} \vartheta^{}_{3}-\varrho^{}_{3} \vartheta^{}_{2} \\
		\varrho^{}_{1} \vartheta^{}_{2}-\varrho^{}_{2} \vartheta^{}_{1} \\
		\varrho^{}_{3} \vartheta^{}_{1}-\varrho^{}_{1} \vartheta^{}_{3}
		\end{array}\right) \\
		{\bf 5}^{}_{\rm s} : \dfrac{\sqrt{6}}{6} \left(\begin{array}{c}
		2 \varrho^{}_{1} \vartheta^{}_{1}-\varrho^{}_{2} \vartheta^{}_{3}-\varrho^{}_{3} \vartheta^{}_{2} \\
		-\sqrt{3} \varrho^{}_{1} \vartheta^{}_{2}-\sqrt{3} \varrho^{}_{2} \vartheta^{}_{1} \\
		\sqrt{6} \varrho^{}_{2} \vartheta^{}_{2} \\
		\sqrt{6} \varrho^{}_{3} \vartheta^{}_{3} \\
		-\sqrt{3} \left(\varrho^{}_{1} \vartheta^{}_{3}+ \varrho^{}_{3} \vartheta^{}_{1} \right)
		\end{array}\right)
		\end{array}\right.$ 
		& 	$\left\{\begin{array}{l}
		{\bf 1}^{}_{\rm s} : \dfrac{\sqrt{3}}{3} \left(\varrho^{}_1 \vartheta^{}_1 + \varrho^{}_2 \vartheta^{}_3 + \varrho^{}_3 \vartheta^{}_2\right)\\
		{\bf 3}^{\prime}_{\rm a} : \dfrac{\sqrt{2}}{2} \left(\begin{array}{c}
		\varrho^{}_{2} \vartheta^{}_{3}-\varrho^{}_{3} \vartheta^{}_{2} \\
		\varrho^{}_{1} \vartheta^{}_{2}-\varrho^{}_{2} \vartheta^{}_{1} \\
		\varrho^{}_{3} \vartheta^{}_{1}-\varrho^{}_{1} \vartheta^{}_{3}
		\end{array}\right) \\
		{\bf 5}^{}_{\rm s} : \dfrac{\sqrt{6}}{6} \left(\begin{array}{c}
		2 \varrho^{}_{1} \vartheta^{}_{1}-\varrho^{}_{2} \vartheta^{}_{3}-\varrho^{}_{3} \vartheta^{}_{2} \\
		\sqrt{6} \varrho^{}_{3} \vartheta^{}_{3} \\
		-\sqrt{3} \left(\varrho^{}_{1} \vartheta^{}_{2}+ \varrho^{}_{2} \vartheta^{}_{1} \right) \\
		-\sqrt{3} \left( \varrho^{}_{1} \vartheta^{}_{3}+ \varrho^{}_{3} \vartheta^{}_{1} \right) \\
		\sqrt{6} \varrho^{}_{2} \vartheta^{}_{2}
		\end{array}\right)
		\end{array}\right.$
		& $\left\{\begin{array}{l}
		{\bf 4} : \dfrac{\sqrt{3}}{3} \left(\begin{array}{c}
		\sqrt{2} \varrho^{}_{2} \vartheta^{}_{1}+\varrho^{}_{3} \vartheta^{}_{2} \\
		-\sqrt{2} \varrho^{}_{1} \vartheta^{}_{2}-\varrho^{}_{3} \vartheta^{}_{3} \\
		-\sqrt{2} \varrho^{}_{1} \vartheta^{}_{3}-\varrho^{}_{2} \vartheta^{}_{2} \\
		\sqrt{2} \varrho^{}_{3} \vartheta^{}_{1}+\varrho^{}_{2} \vartheta^{}_{3}
		\end{array}\right) \\
		{\bf 5} : \dfrac{\sqrt{3}}{3} \left(\begin{array}{c}
		\sqrt{3} \varrho^{}_{1} \vartheta^{}_{1} \\
		\varrho^{}_{2} \vartheta^{}_{1}-\sqrt{2} \varrho^{}_{3} \vartheta^{}_{2} \\
		\varrho^{}_{1} \vartheta^{}_{2}-\sqrt{2} \varrho^{}_{3} \vartheta^{}_{3} \\
		\varrho^{}_{1} \vartheta^{}_{3}-\sqrt{2} \varrho^{}_{2} \vartheta^{}_{2} \\
		\varrho^{}_{3} \vartheta^{}_{1}-\sqrt{2} \varrho^{}_{2} \vartheta^{}_{3}
		\end{array}\right)
		\end{array}\right.$
		\vspace{0.5cm}
	\end{tabular}
\end{small}
%\vspace{-1cm}
\renewcommand\arraystretch{1.2}
	\begin{tabular}{L{8.4cm}L{8.4cm}}
		\hspace{2mm}${\bf 4} \otimes {\bf 4} = {\bf 1}^{}_{\rm s} \oplus {\bf 3}^{}_{\rm a} \oplus {\bf 3}^{\prime}_{\rm a} \oplus {\bf 4}^{}_{\rm s} \oplus {\bf 5}^{}_{\rm s}$ & \hspace{2mm}$\widehat{\bf 4} \otimes \widehat{\bf 4} = {\bf 1}^{}_{\rm a} \oplus {\bf 3}^{}_{\rm s} \oplus {\bf 3}^{\prime}_{\rm s} \oplus {\bf 4}^{}_{\rm s} \oplus {\bf 5 }^{}_{\rm s} $ \\
		$\left\{\begin{array}{l}
		{\bf 1}^{}_{\rm s}: \dfrac{1}{2} \left(\varrho^{}_{1} \vartheta^{}_{4}+\varrho^{}_{2} \vartheta^{}_{3}+\varrho^{}_{3} \vartheta^{}_{2}+\varrho^{}_{4} \vartheta^{}_{1}\right) \\
		{\bf 3}^{}_{\rm a} : \dfrac{1}{2} \left(\begin{array}{c}
		-\varrho^{}_{1} \vartheta^{}_{4}+\varrho^{}_{2} \vartheta^{}_{3}-\varrho^{}_{3} \vartheta^{}_{2}+\varrho^{}_{4} \vartheta^{}_{1} \\
		\sqrt{2} \left( \varrho^{}_{2} \vartheta^{}_{4}- \varrho^{}_{4} \vartheta^{}_{2} \right) \\
		\sqrt{2} \left( \varrho^{}_{1} \vartheta^{}_{3}- \varrho^{}_{3} \vartheta^{}_{1} \right)
		\end{array}\right) \\
		{\bf 3}^{\prime}_{\rm a} : \dfrac{1}{2} \left(\begin{array}{c}
		\varrho^{}_{1} \vartheta^{}_{4}+\varrho^{}_{2} \vartheta^{}_{3}-\varrho^{}_{3} \vartheta^{}_{2}-\varrho^{}_{4} \vartheta^{}_{1} \\
		\sqrt{2} \left( \varrho^{}_{3} \vartheta^{}_{4}- \varrho^{}_{4} \vartheta^{}_{3} \right) \\
		\sqrt{2} \left( \varrho^{}_{1} \vartheta^{}_{2}- \varrho^{}_{2} \vartheta^{}_{1} \right)
		\end{array}\right) \\
		{\bf 4}^{}_{\rm s} : \dfrac{\sqrt{3}}{3} \left(\begin{array}{c}
		\varrho^{}_{2} \vartheta^{}_{4}+\varrho^{}_{3} \vartheta^{}_{3}+\varrho^{}_{4} \vartheta^{}_{2} \\
		\varrho^{}_{1} \vartheta^{}_{1}+\varrho^{}_{3} \vartheta^{}_{4}+\varrho^{}_{4} \vartheta^{}_{3} \\
		\varrho^{}_{1} \vartheta^{}_{2}+\varrho^{}_{2} \vartheta^{}_{1}+\varrho^{}_{4} \vartheta^{}_{4} \\
		\varrho^{}_{1} \vartheta^{}_{3}+\varrho^{}_{2} \vartheta^{}_{2}+\varrho^{}_{3} \vartheta^{}_{1}
		\end{array}\right)\\
		{\bf 5}^{}_{\rm s} : \dfrac{\sqrt{3}}{6} \left(\begin{array}{c}
		\sqrt{3} \left( \varrho^{}_{1} \vartheta^{}_{4}- \varrho^{}_{2} \vartheta^{}_{3}- \varrho^{}_{3} \vartheta^{}_{2}+ \varrho^{}_{4} \vartheta^{}_{1}  \right) \\
		-\sqrt{2} \left(\varrho^{}_{2} \vartheta^{}_{4}- 2  \varrho^{}_{3} \vartheta^{}_{3}+ \varrho^{}_{4} \vartheta^{}_{2} \right) \\
		- \sqrt{2} \left( 2\varrho^{}_{1} \vartheta^{}_{1}- \varrho^{}_{3} \vartheta^{}_{4}- \varrho^{}_{4} \vartheta^{}_{3} \right) \\
		\sqrt{2} \left( \varrho^{}_{1} \vartheta^{}_{2}+ \varrho^{}_{2} \vartheta^{}_{1}-2 \varrho^{}_{4} \vartheta^{}_{4} \right)  \\
		-\sqrt{2} \left( \varrho^{}_{1} \vartheta^{}_{3} - 2 \varrho^{}_{2} \vartheta^{}_{2}+ \varrho^{}_{3} \vartheta^{}_{1}\right)
		\end{array}\right)
		\end{array}\right.$
		& $\left\{\begin{array}{l}
		{\bf 1}^{}_{\rm a} : \dfrac{1}{2} \left(\varrho^{}_1 \vartheta^{}_4 + \varrho^{}_2 \vartheta^{}_3 - \varrho^{}_3 \vartheta^{}_2 - \varrho^{}_4 \vartheta^{}_1 \right) \\
		{\bf 3}^{}_{\rm s} : -\dfrac{\sqrt{5}}{10} \left(\begin{array}{r}
		3 \varrho_{1}^{} \vartheta_{4}^{}+\varrho_{2}^{} \vartheta_{3}^{}+\varrho_{3}^{} \vartheta_{2}^{}+3 \varrho_{4}^{} \vartheta_{1}^{} \\
		\sqrt{2}\left(\sqrt{3} \varrho_{2}^{} \vartheta_{4}^{}-2 \varrho_{3}^{} \vartheta_{3}^{}+\sqrt{3} \varrho_{4}^{} \vartheta_{2}^{}\right) \\
		\sqrt{2}\left(\sqrt{3} \varrho_{1}^{} \vartheta_{3}^{}+2 \varrho_{2}^{} \vartheta_{2}^{}+\sqrt{3} \varrho_{3}^{} \vartheta_{1}^{}\right)
		\end{array}\right) \\
		{\bf 3}^{\prime}_{\rm s} : -\dfrac{\sqrt{5}}{10} \left(\begin{array}{c}
		\varrho_{1}^{} \vartheta_{4}^{}-3 \varrho_{2}^{} \vartheta_{3}^{}-3 \varrho_{3}^{} \vartheta_{2}^{}+\varrho_{4}^{} \vartheta_{1}^{} \\
		\sqrt{2}\left(2 \varrho_{1}^{} \vartheta_{1}^{}-\sqrt{3}\varrho_{3}^{} \vartheta_{4}^{}-\sqrt{3}\varrho_{4}^{} \vartheta_{3}^{}\right) \\
		\sqrt{2}\left(\sqrt{3} \varrho_{1}^{} \vartheta_{2}^{}+\sqrt{3} \varrho_{2}^{} \vartheta_{1}^{}-2 \varrho_{4}^{} \vartheta_{4}^{}\right)
		\end{array}\right)\\
		{\bf 4}^{}_{\rm s} : \dfrac{\sqrt{5}}{5} \left(\begin{array}{c}
		\varrho_{2}^{} \vartheta_{4}^{}+\sqrt{3} \varrho_{3}^{} \vartheta_{3}^{}+\varrho_{4}^{} \vartheta_{2}^{} \\
		-\sqrt{3} \varrho_{1}^{} \vartheta_{1}^{}-\varrho_{3}^{} \vartheta_{4}^{}-\varrho_{4}^{} \vartheta_{3}^{} \\
		-\varrho_{1}^{} \vartheta_{2}^{}-\varrho_{2}^{} \vartheta_{1}^{}-\sqrt{3} \varrho_{4}^{} \vartheta_{4}^{} \\
		-\varrho_{1}^{} \vartheta_{3}^{}+\sqrt{3} \varrho_{2}^{} \vartheta_{2}^{}-\varrho_{3}^{} \vartheta_{1}^{}
		\end{array}\right) \\
		{\bf 5}^{}_{\rm a} :  \dfrac{1}{2} \left(\begin{array}{c}
		\varrho_{1}^{} \vartheta_{4}^{}-\varrho_{2}^{} \vartheta_{3}^{}+\varrho_{3}^{} \vartheta_{2}^{}-\varrho_{4}^{} \vartheta_{1}^{} \\
		-\sqrt{2}\left(\varrho_{2}^{} \vartheta_{4}^{}-\varrho_{4}^{} \vartheta_{2}^{}\right) \\
		-\sqrt{2}\left(\varrho_{3}^{} \vartheta_{4}^{}-\varrho_{4}^{} \vartheta_{3}^{}\right) \\
		\sqrt{2}\left(\varrho_{1}^{} \vartheta_{2}^{}-\varrho_{2}^{} \vartheta_{1}^{}\right) \\
		-\sqrt{2}\left(\varrho_{1}^{} \vartheta_{3}^{}-\varrho_{3}^{} \vartheta_{1}^{}\right)
		\end{array}\right)
		\end{array}\right.$
		\vspace{0.5cm}
	\end{tabular}\\
	
		\renewcommand\arraystretch{1.23}
	\begin{tabular}{L{17cm}}
		${\bf 5} \otimes {\bf 5} = {\bf 1}^{}_{\rm s} \oplus {\bf 3}^{}_{\rm a} \oplus {\bf 3}^{\prime}_{\rm a} \oplus {\bf 4}^{}_{\rm s} \oplus {\bf 4}^{}_{\rm a} \oplus {\bf 5}^{}_{\rm s,1} \oplus {\bf 5}^{}_{\rm s,2} $\\
		$\left\{\begin{array}{l}
		{\bf 1}^{}_{\rm s} : \dfrac{\sqrt{5}}{5} \left(\varrho^{}_{1} \vartheta^{}_{1}+\varrho^{}_{2} \vartheta^{}_{5}+\varrho^{}_{3} \vartheta^{}_{4}+\varrho^{}_{4} \vartheta^{}_{3}+\varrho^{}_{5} \vartheta^{}_{2}\right) \\
		{\bf 3}^{}_{\rm a} : \dfrac{\sqrt{10}}{10} \left(\begin{array}{c}
		\varrho^{}_{2} \vartheta^{}_{5}+2 \varrho^{}_{3} \vartheta^{}_{4}-2 \varrho^{}_{4} \vartheta^{}_{3}-\varrho^{}_{5} \vartheta^{}_{2} \\
		-\sqrt{3} \varrho^{}_{1} \vartheta^{}_{2}+\sqrt{3} \varrho^{}_{2} \vartheta^{}_{1}+\sqrt{2} \varrho^{}_{3} \vartheta^{}_{5}-\sqrt{2} \varrho^{}_{5} \vartheta^{}_{3} \\
		\sqrt{3} \varrho^{}_{1} \vartheta^{}_{5}+\sqrt{2} \varrho^{}_{2} \vartheta^{}_{4}-\sqrt{2} \varrho^{}_{4} \vartheta^{}_{2}-\sqrt{3} \varrho^{}_{5} \vartheta^{}_{1}
		\end{array}\right)  \\
		{\bf 3}^{\prime}_{\rm a} : \dfrac{\sqrt{10}}{10} \left(\begin{array}{c}
		2 \varrho^{}_{2} \vartheta^{}_{5}-\varrho^{}_{3} \vartheta^{}_{4}+\varrho^{}_{4} \vartheta^{}_{3}-2 \varrho^{}_{5} \vartheta^{}_{2} \\
		\sqrt{3} \varrho^{}_{1} \vartheta^{}_{3}-\sqrt{3} \varrho^{}_{3} \vartheta^{}_{1}+\sqrt{2} \varrho^{}_{4} \vartheta^{}_{5}-\sqrt{2} \varrho^{}_{5} \vartheta^{}_{4} \\
		-\sqrt{3} \varrho^{}_{1} \vartheta^{}_{4}+\sqrt{2} \varrho^{}_{2} \vartheta^{}_{3}-\sqrt{2} \varrho^{}_{3} \vartheta^{}_{2}+\sqrt{3} \varrho^{}_{4} \vartheta^{}_{1}
		\end{array}\right) \\
		{\bf 4}^{}_{\rm s} : \dfrac{\sqrt{30}}{30} \left(\begin{array}{l}
		\sqrt{6} \varrho^{}_{1} \vartheta^{}_{2}+\sqrt{6} \varrho^{}_{2} \vartheta^{}_{1}- \varrho^{}_{3} \vartheta^{}_{5}+4  \varrho^{}_{4} \vartheta^{}_{4}- \varrho^{}_{5} \vartheta^{}_{3} \\
		\sqrt{6} \varrho^{}_{1} \vartheta^{}_{3}+4  \varrho^{}_{2} \vartheta^{}_{2}+\sqrt{6} \varrho^{}_{3} \vartheta^{}_{1}- \varrho^{}_{4} \vartheta^{}_{5}- \varrho^{}_{5} \vartheta^{}_{4} \\
		\sqrt{6} \varrho^{}_{1} \vartheta^{}_{4}- \varrho^{}_{2} \vartheta^{}_{3}- \varrho^{}_{3} \vartheta^{}_{2}+\sqrt{6} \varrho^{}_{4} \vartheta^{}_{1}+4  \varrho^{}_{5} \vartheta^{}_{5} \\
		\sqrt{6} \varrho^{}_{1} \vartheta^{}_{5}- \varrho^{}_{2} \vartheta^{}_{4}+4  \varrho^{}_{3} \vartheta^{}_{3}- \varrho^{}_{4} \vartheta^{}_{2}+\sqrt{6} \varrho^{}_{5} \vartheta^{}_{1}
		\end{array}\right)\\
		{\bf 4}^{}_{\rm a} : \dfrac{\sqrt{10}}{10} \left(\begin{array}{c}
		\sqrt{2} \varrho^{}_{1} \vartheta^{}_{2}-\sqrt{2} \varrho^{}_{2} \vartheta^{}_{1}+\sqrt{3} \varrho^{}_{3} \vartheta^{}_{5}-\sqrt{3} \varrho^{}_{5} \vartheta^{}_{3} \\
		-\sqrt{2} \varrho^{}_{1} \vartheta^{}_{3}+\sqrt{2} \varrho^{}_{3} \vartheta^{}_{1}+\sqrt{3} \varrho^{}_{4} \vartheta^{}_{5}-\sqrt{3} \varrho^{}_{5} \vartheta^{}_{4} \\
		-\sqrt{2} \varrho^{}_{1} \vartheta^{}_{4}-\sqrt{3} \varrho^{}_{2} \vartheta^{}_{3}+\sqrt{3} \varrho^{}_{3} \vartheta^{}_{2}+\sqrt{2} \varrho^{}_{4} \vartheta^{}_{1} \\
		\sqrt{2} \varrho^{}_{1} \vartheta^{}_{5}-\sqrt{3} \varrho^{}_{2} \vartheta^{}_{4}+\sqrt{3} \varrho^{}_{4} \vartheta^{}_{2}-\sqrt{2} \varrho^{}_{5} \vartheta^{}_{1}
		\end{array}\right) \\
		{\bf 5}^{}_{\rm s,1} : \dfrac{\sqrt{14}}{14} \left(\begin{array}{c}
		2 \varrho^{}_{1} \vartheta^{}_{1}+\varrho^{}_{2} \vartheta^{}_{5}-2 \varrho^{}_{3} \vartheta^{}_{4}-2 \varrho^{}_{4} \vartheta^{}_{3}+\varrho^{}_{5} \vartheta^{}_{2} \\
		\varrho^{}_{1} \vartheta^{}_{2}+\varrho^{}_{2} \vartheta^{}_{1}+\sqrt{6} \varrho^{}_{3} \vartheta^{}_{5}+\sqrt{6} \varrho^{}_{5} \vartheta^{}_{3} \\
		-2 \varrho^{}_{1} \vartheta^{}_{3}+\sqrt{6} \varrho^{}_{2} \vartheta^{}_{2}-2 \varrho^{}_{3} \vartheta^{}_{1} \\
		-2 \varrho^{}_{1} \vartheta^{}_{4}-2 \varrho^{}_{4} \vartheta^{}_{1}+\sqrt{6} \varrho^{}_{5} \vartheta^{}_{5} \\
		\varrho^{}_{1} \vartheta^{}_{5}+\sqrt{6} \varrho^{}_{2} \vartheta^{}_{4}+\sqrt{6} \varrho^{}_{4} \vartheta^{}_{2}+\varrho^{}_{5} \vartheta^{}_{1}
		\end{array}\right)   \\
		{\bf 5}^{}_{\rm s,2} : \dfrac{\sqrt{14}}{14} \left(\begin{array}{c}
		2 \varrho_{1} \vartheta_{1}-2 \varrho_{2} \vartheta_{5}+\varrho_{3} \vartheta_{4}+\varrho_{4} \vartheta_{3}-2 \varrho_{5} \vartheta_{2} \\
		-2 \varrho_{1} \vartheta_{2}-2 \varrho_{2} \vartheta_{1}+\sqrt{6} \varrho_{4} \vartheta_{4} \\
		\varrho_{1} \vartheta_{3}+\varrho_{3} \vartheta_{1}+\sqrt{6} \varrho_{4} \vartheta_{5}+\sqrt{6} \varrho_{5} \vartheta_{4} \\
		\varrho_{1} \vartheta_{4}+\sqrt{6} \varrho_{2} \vartheta_{3}+\sqrt{6} \varrho_{3} \vartheta_{2}+\varrho_{4} \vartheta_{1} \\
		-2 \varrho_{1} \vartheta_{5}+\sqrt{6} \varrho_{3} \vartheta_{3}-2 \varrho_{5} \vartheta_{1}
		\end{array}\right)
		\end{array}\right.$
		\vspace{0.6cm}
	\end{tabular}

\bibliographystyle{utcaps_mod}
\bibliography{linear_a5}

\end{document}